\documentclass[sigplan,screen,10pt]{acmart} 

\acmYear{2026}\copyrightyear{2026}
\acmConference[EUROSYS '26]{European Conference on Computer Systems}{April 27--30, 2026}{Edinburgh, Scotland UK}
\acmBooktitle{European Conference on Computer Systems (EUROSYS '26), April 27--30, 2026, Edinburgh, Scotland UK}
\acmDOI{10.1145/3767295.3769384}
\acmISBN{979-8-4007-2212-7/26/04}

\title[ASIC-based Compression Accelerators for Storage Systems]%
{ASIC-based Compression Accelerators for Storage Systems: Design, Placement, and Profiling Insights}

\usepackage{multirow}
\usepackage{soul}
\usepackage{mdframed}
\usepackage{amsmath}
\usepackage{graphicx}
\usepackage{hyperref}
\usepackage{listings}
\usepackage[htt]{hyphenat}
\usepackage{subcaption}
\usepackage{color}
\usepackage{array}
\usepackage{enumitem}
\usepackage{pifont}
\usepackage{xurl}
\usepackage[most]{tcolorbox}

\usepackage{pifont}      
\usepackage{tikz}        
\usetikzlibrary{positioning}

\newcommand{\halfcheck}{%
  \tikz[baseline=-0.2ex,inner sep=0,outer sep=0]{%
    \node[inner sep=0,outer sep=0] (C) {\ding{51}};
    \draw[line width=0.1em]
      ([shift={(-0.8ex,-0.2ex)}]C.north east) -- ++(0.25ex,-0.9ex);
  }%
}

\AtBeginDocument{%
  }

\begin{document}

\author{Tao Lu}
\affiliation{%
  \institution{DapuStor}
  \city{Shenzhen}
  \country{China}
}

\author{Jiapin Wang}
\affiliation{%
  \institution{DapuStor}
  \city{Shenzhen}
  \country{China}
}

\author{Yelin Shan}
\affiliation{%
  \institution{DapuStor}
  \city{Shenzhen}
  \country{China}
}

\author{Xiangping Zhang}
\affiliation{%
  \institution{DapuStor}
  \city{Shenzhen}
  \country{China}
}

\author{Xiang Chen}
\authornote{Xiang Chen is the corresponding author (chenxiang@dapustor.com). He is also affiliated with HUST during the research.}
\affiliation{%
  \institution{DapuStor}
  \city{Shenzhen}
  \country{China}
}

\begin{abstract}
Lossless compression imposes significant computational overhead on datacenters when performed on CPUs. Hardware compression and decompression processing units (CDPUs) can alleviate this overhead, but optimal algorithm selection, microarchitectural design, and system-level placement of CDPUs are still not well understood. We present the design of an ASIC-based in-storage CDPU and provide a comprehensive end-to-end evaluation against two leading ASIC accelerators, Intel QAT 8970 and QAT 4xxx. The evaluation spans three dominant CDPU placement regimes: peripheral, on-chip, and in-storage. Our results reveal: (\textbf{i}) acute sensitivity of throughput and latency to CDPU placement and interconnection, (\textbf{ii}) strong correlation between compression efficiency and data patterns/layouts, (\textbf{iii}) placement-driven divergences between microbenchmark gains and real-application speedups, (\textbf{iv}) discrepancies between module and system-level power efficiency, and (\textbf{v}) scalability and multi-tenant interference issues of various CDPUs. These findings motivate a placement-aware, cross-layer rethinking of hardware (de)compression for hyperscale storage infrastructures.
\end{abstract}

\ccsdesc[500]{Information systems~Storage architectures}

\keywords{ASIC Compression Accelerator, Storage System}

\setcopyright{cc}
\setcctype[4.0]{by-nc-nd}

\maketitle

\section{Introduction}

Data compression enables hyperscale storage systems to significantly reduce storage footprint, energy consumption, and transmission latency~\cite{ziv1977universal,huffman1952method,deutsch1996rfc1951,b1_ibmz15_2020}. Compression algorithms such as Ziv-Lempel~\cite{ziv1977universal} and Huffman~\cite{huffman1952method} underpin standards like Deflate~\cite{deutsch1996rfc1951}, now ubiquitously deployed from filesystems (e.g., ZFS~\cite{qzfs2019}, Btrfs~\cite{rodeh2013btrfs}) and virtualization platforms (e.g., MicroVMs~\cite{lazarev2024sabre}) to cloud systems (e.g., Microsoft Zipline~\cite{microsoft2019zipline} and Google datacenter~\cite{karandikar2023cdpu}). 

Dictionary and entropy coding impose high computational overhead during compression. Algorithm design typically trades off computational complexity against compression ratio\footnote{Compression ratio is defined as the size of compressed data divided by the size of original data; smaller values indicate better compression.}. Consequently, there exist both high-compression ``standard'' algorithms such as Deflate~\cite{deutsch1996rfc1951} and Zstd~\cite{zstd} and low-compression, lightweight algorithms like Snappy~\cite{snappy} and LZ4~\cite{lz4}, with median compression ratio differences reaching 20\% (e.g., 40\% vs 60\% as shown in Figure~\ref{fig:DPZip_cr}). Although 95\% of compressed bytes in Google's fleet rely on lightweight algorithms such as Snappy~\cite{snappy}, prioritizing CPU offloading over optimal compression ratios, software (de)compression still accounts for 2.9\% of total fleet CPU cycles, and as much as 10--50\% in critical services~\cite{karandikar2023cdpu}.

\begin{figure}[t]
    \centering
    \includegraphics[width=8cm]{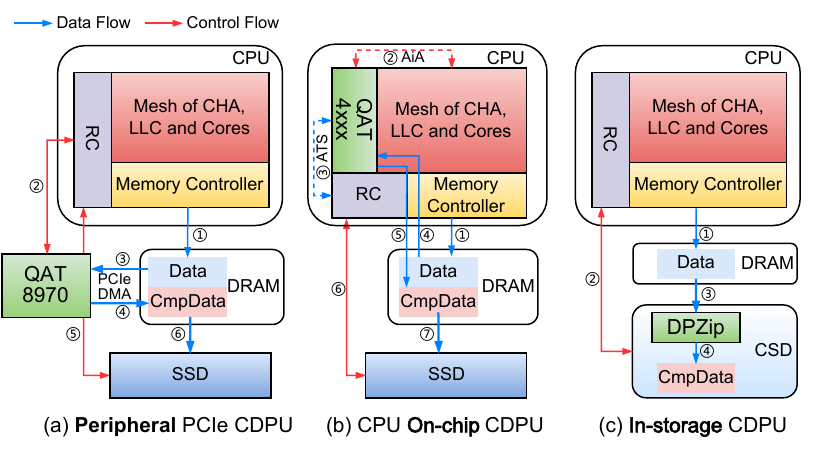}
    \vspace{-1em}
    \caption{Various hardware CDPU accelerator architectures. RC = PCIe root complex; CHA = Caching and Home Agent (LLC tag lookup and cache coherence).}
    \label{fig:architecture}
\end{figure}

Hardware compression and decompression processing units (CDPUs) offload CPU overhead from standard compression algorithms while maintaining compression ratios. CDPU architectures have evolved from FPGA-based implementations~\cite{abdelfattah2014Gziponachip,fowers2015scalable} and PCIe cards~\cite{qzfs2019,gao2022metazip} to chiplet integration within CPU processors~\cite{b1_ibmz15_2020,yuan2024intel} and SSD controllers~\cite{b5_csd2000,wang2025reviving}. Industry deployments including Microsoft's Zipline~\cite{microsoft2019zipline}, Intel QAT~\cite{website:qat,yuan2024intel}, and ScaleFlux's CSD~\cite{scaleflux_csd2000} represent three dominant placement strategies (Figure~\ref{fig:architecture}): \textbf{(1) peripheral} PCIe-attached cards (QAT 8970~\cite{website:qat}) with standard PCIe interfaces; \textbf{(2) on-chip} ASIC chiplets (QAT 4xxx~\cite{yuan2024intel}) integrated on CPU dies using cache-coherent mesh~\cite{yuan2024intel,lou2025dynamic} or DMA interfaces~\cite{b1_ibmz15_2020} to eliminate PCIe latency; and \textbf{(3) in-storage} CDPUs within SSD controllers performing IO-path compression to minimize data movement.
Modern CDPUs achieve sub-$10\mu\mathrm{s}$ latencies and throughputs exceeding $13.8\mathrm{GB/s}$, delivering over 30$\times$ performance improvements versus single-threaded CPU compression~\cite{b1_ibmz15_2020}. 

Despite significant progress in CDPU evolution, key research gaps persist. Prior work predominantly targets single CDPUs~\cite{qzfs2019, b1_ibmz15_2020, yuan2024intel} or relies on simulation and modeling~\cite{karandikar2023cdpu, beezip2024asplos}, lacking side-by-side evaluations of multiple genuine ASIC-based CDPUs. 
For instance, on-chip CDPUs are widely assumed to advance beyond peripheral designs and deliver substantial performance gains. However, our evaluation reveals that on-chip CDPUs provide no bandwidth improvements and primarily reduce latency, which is translated into higher end-to-end application throughput. This nuanced finding, with direct implications for future system designs, has not been previously discussed.
Moreover, while peripheral~\cite{qzfs2019, amd2023maxlinear} and on-chip~\cite{b1_ibmz15_2020, karandikar2023cdpu, yuan2024intel} CDPUs have been well-explored, in-storage CDPUs remain underexplored, despite their strong potential for system-level improvements in performance and energy efficiency~\cite{snia_csd_v1, b5_csd2000, chen2024hacsd, wang2025reviving}.

Furthermore, average throughput and latency are insufficient to fully assess the impact of CDPU integration, as they overlook critical factors such as host CPU utilization, CDPU-PCIe-memory hierarchy interactions, and architectural effects on system scalability. These elements introduce complex interdependencies that simplistic ``faster accelerator'' metrics fail to capture. Consequently, a rigorous evaluation framework should comprehensively analyze system-wide interactions, considering these dependencies rather than focusing solely on isolated accelerator performance gains.

In summary, existing studies have not addressed \textbf{(i)} the design trade‑offs of CDPUs in resource‑constrained environments such as SSD controllers, \textbf{(ii)} the impact of placing CDPUs at different locations in the storage stack on systems and applications, and \textbf{(iii)} the scalability and performance isolation of CDPUs in multi‑tenant environments. Therefore, a holistic, empirical approach that considers microarchitectural features to evaluate CDPU placement strategies and runtime parameters is essential for rigorously quantifying CDPU trade‑offs and benefits in modern storage systems.

To address the above issues, we present DPZip, the first disclosure of an in‑storage ASIC compression accelerator embedded in commercial PCIe 5.0 SSDs. Occupying about 5\% of the SSD controller die area and running at 1GHz, DPZip processes 8 bytes per cycle via a dynamic Huffman engine and parallel pipeline. On SSD-friendly 4KB pages, it surpasses existing in‑storage FPGA~\cite{b5_csd2000} and Intel QAT~\cite{qzfs2019,yuan2024intel} hardware solutions in performance while delivering comparable compression ratios. 
DPZip unlocks a new point in the spectrum of accelerator placements, complementing PCIe‑peripheral and on‑chip solutions~\cite{b1_ibmz15_2020,karandikar2023cdpu,yuan2024intel}.

We further conduct a comprehensive, workload-driven analysis of CDPU placement effects on end-to-end storage performance using multiple real ASIC CDPUs. Our results demonstrate that compression-induced latency limits application throughput, emphasizing the necessity for workload-aware deployment strategies. Our cross-device and cross-workload comparisons reveal key performance patterns and behaviors, offering valuable insights for root-cause diagnosis and storage system optimizations.

Finally, we systematically evaluate CDPU's performance scalability across multi-threading and multi-device configurations, as well as VF performance isolation in multi-tenant environments. With Intel QAT~\cite{website:qat,yuan2024intel}, we demonstrate that conventional SR-IOV mechanisms fail to prevent detrimental ``noisy neighbor'' effects. CDPU-internal fair resource allocation mechanisms are essential for robust multi-tenant operation. Through comprehensive quantification across varying device counts and placements, this study delivers actionable design guidance for cost-effective, scalable multi-tenant accelerator architectures with performance isolation.

In summary, our core contributions include:
\vspace{-0.5em}
\begin{itemize}[leftmargin=*]
    \item \textbf{DPZip accelerator and DP-CSD device.} DPZip presents the first disclosure of in-storage ASIC CDPU hardware algorithm design—including a resource-efficient LZ77 encoder and an adaptive dynamic Huffman encoder. DP-CSD presents the internals of a compression-enabled SSD, especially the integration of the CDPU in resource-constrained SSD environments, as well as the sophisticated flash translation layer (FTL) data management. 
    
    \item The first \textbf{genuine ASIC-based CDPU comparative study} of accelerator placement in real systems running real-world database and filesystem applications. Our study fully covers mainstream CDPU placement options. Specifically, we benchmark in-storage DPZip CDPU against state-of-the-art peripheral (e.g., QAT 8970) and on-chip (e.g., QAT 4xxx) solutions, analyzing performance, compression ratio, and energy efficiency. 

    \item The first study assessing \textbf{CDPU scalability and performance isolation} in multi-tenant SR-IOV~\cite{dong2012high} virtualization environments. Specifically, each CDPU device is partitioned into 24 Virtual Functions(VFs) and individually mapped to 24 virtual machines(VMs) to evaluate performance and consistency.

    \item Open-sourced evaluation artifacts including datasets and tools at \url{https://github.com/Emilio597/ASIC-CDPUs}.
    
\end{itemize}

The rest of this paper is organized as follows. Section~\ref{backgrd} reviews background, related work, and presents our research motivation. Sections~\ref{design} and \ref{dpcsd} describe the design and implementation of DPZip and its integration into DP-CSD. Section~\ref{evaluation} presents profiling results, while Section~\ref{discuss} discusses broader implications and future work.

\section{Background and Motivation}\label{backgrd}
\subsection{Hyperscale Storage Demands Data Compression}
Modern computing infrastructures face an exponential increase in data, making effective compression indispensable for hyperscale storage systems~\cite{fb_zstandard_2018,karandikar2023cdpu}. Cloud platforms achieve noteworthy benefits by using 2-4$\times$ compression ratios that yield six-figure annual savings per petabyte~\cite{b1_ibmz15_2020}. 
Enterprise solutions such as IBM zEDC~\cite{ibm_zedc}, RedHat VDO~\cite{horn2018vdo}, Meta's compression service~\cite{fb_zstandard_2018} have integrated advanced compression techniques to significantly reduce storage footprints. Furthermore, modern filesystems like ZFS~\cite{qzfs2019} and Btrfs~\cite{rodeh2013btrfs} leverage compression for improved data management, while systems such as RocksDB~\cite{rocksdb_github} adopt hybrid compression strategies to balance performance and efficiency. Overall, these developments confirm that advanced compression is crucial for delivering cost-effective, scalable hyperscale storage solutions.

\subsection{Compression: A Timeless, Complex Technique}
Lossless compression relies primarily on two paradigms: dictionary compression and entropy coding. Dictionary-based methods, exemplified by LZ77~\cite{ziv1977universal}, efficiently capture data redundancy using dynamic sliding windows and offset-length-literal tuple encoding. In parallel, entropy coding has progressed from traditional approaches, such as Huffman coding~\cite{huffman1952method}, to more sophisticated schemes like Asymmetric Numeral Systems (ANS)~\cite{Duda2013} and Finite State Entropy (FSE)~\cite{collet2013fse}, further enhancing compression performance.

\begin{figure}[t]
    \centering
    \includegraphics[width=8.5cm]{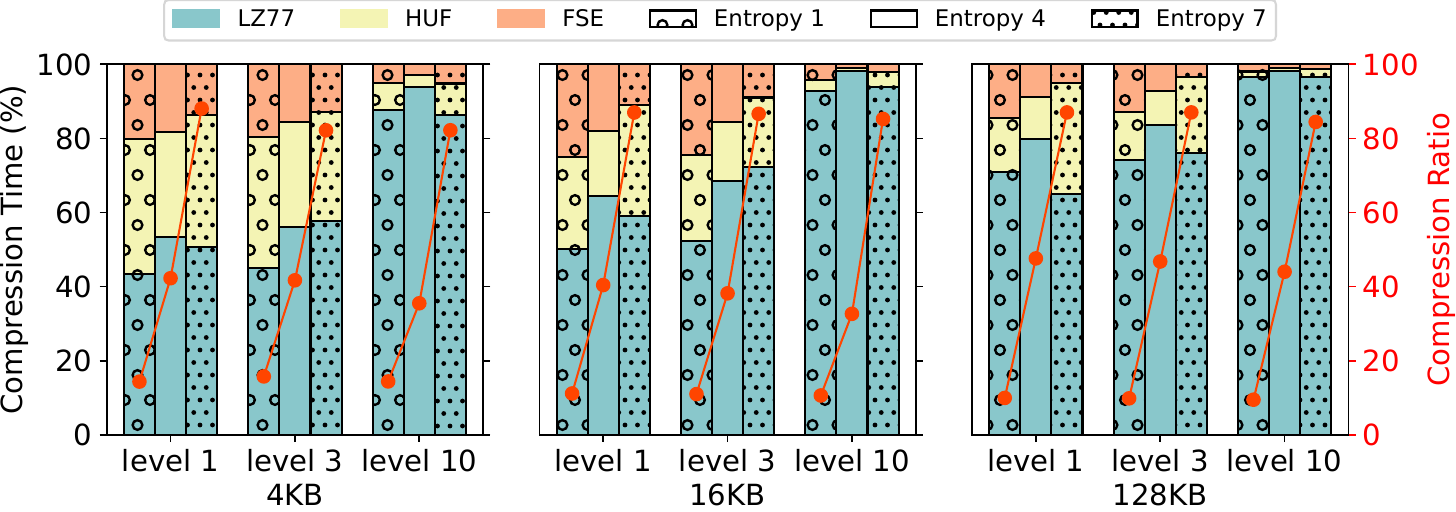}
    \vspace{-2em}
    \caption{Zstd execution time breakdown demonstrates the complicated interplay among compression granularities, levels, and data patterns (entropy values).}
    \label{fig:Zstdbreakdown}
\end{figure}

Modern compression algorithms such as Zstd unify diverse paradigms. Our analysis (Figure~\ref{fig:Zstdbreakdown}) reveals that LZ77 operations dominate computational costs, especially at higher compression levels. Critical parameters—including chunk size (4KB–128KB), compression level, and Shannon entropy\footnote{The Shannon entropy $H(X)$ of a dataset, in the context of data compression, represents the theoretical minimum average number of bits needed to encode each symbol, and is given by $H(X) = - \sum\limits_{x \in \mathcal{X}} p(x)\,\log_{2} p(x)$ where $p(x)$ is the probability of each symbol $x$.}, jointly determine both effectiveness and efficiency. Compression level primarily controls LZ77's pattern search depth, whereas Shannon entropy measures inherent data compressibility. Notably, the computational impact of entropy coding stages (Huffman and FSE) decreases with higher compression levels, but varies nonlinearly with data randomness. This complex parameter space demands rigorous optimization to balance computational overhead, speed, and compression ratio, as underscored in a recent study~\cite{karandikar2023cdpu}.

\subsection{CDPU: Hardware Acceleration for Efficiency}
\textbf{Accelerators in storage contexts.} Hardware acceleration provides substantial performance benefits in storage systems, with early programmable prototypes like BlueDBM and SmartSSD demonstrating significant improvements~\cite{Jun2015ISCA-BlueDBM,Salamat2021FPGA-NASCENT,Fakhry2023Array-Review}. Modern SSDs leverage integrated accelerators for diverse applications: in-storage analytics and forensic decoding~\cite{Satyanarayanan2022Micro-SilentWitness}, AI pipeline optimization through distributed pre/post-processing~\cite{Okafor2024JETC-Fuse}, native data services including key-value operations~\cite{Zheng2022LCTES-ISKEVA}, autonomous stream management~\cite{StreamCSD}, and retrieval augmented generation acceleration~\cite{mahapatra2025storage}. These computational storage solutions consistently achieve superior performance for diverse workloads~\cite{Do2020TOS-Newport,Heydarigorji2022TECS-CSD,Fakhry2023Array-Review}.

\textbf{CDPU accelerators.} Traditional data (de)compression algorithms, such as Deflate~\cite{deutsch1996rfc1951}, used in Gzip and ZLIB~\cite{gzip, zlib}, offer relatively high compression ratios but are limited to throughput in the hundreds of MB/s range because of their algorithmic complexity.
While lightweight alternatives such as LZ4~\cite{lz4} and Snappy~\cite{snappy} offer improved speed, they sacrifice compression effectiveness. Even newer solutions like Zstd~\cite{zstd} demand substantial CPU resources. Recent studies indicate compression workloads consume 2.9-50\% of CPU cycles~\cite{kanev2015profiling,sriraman2020accelerometer,gonzalez2023profiling}, with hyperscale environments processing up to 95\% of data using lightweight compression methods~\cite{karandikar2023cdpu,li2023more}.
Hardware CDPU has become a promising solution to enable favorable compression ratio without burdening CPUs, manifesting in three distinct architectural approaches as Figure~\ref{fig:architecture} illustrates. 
\begin{itemize}[leftmargin=*]
    \item \textbf{Peripheral} accelerators, including FPGAs~\cite{chen2021fpga,fowers2015scalable,qiao2018high}and ASICs~\cite{website:qat,qzfs2019}, offloading compression from host CPU to PCIe-attached hardware CDPUs to simultaneously achieve high performance and favorable compression ratio. 
    \item \textbf{On-chip} accelerators, such as IBM's NXU~\cite{b1_ibmz15_2020} and Intel's CPU-integrated QAT~\cite{yuan2024intel}, optimize data movement through direct memory subsystem integration.
    \item \textbf{In-storage} accelerators, exemplified by ScaleFlux's CSD series~\cite{scaleflux_csd2000}, execute compression in the SSD IO path, eliminating host-CDPU data movement for compression.
\end{itemize}

CDPU architectures reflect inherent trade-offs among performance, flexibility, and resource efficiency. Peripheral accelerators~\cite{website:qat, amd2023maxlinear} offload compression with high efficiency, but suffer data movement bottlenecks over PCIe and add system complexity~\cite{qzfs2019}. In contrast, on-chip accelerators such as IBM's NXU~\cite{b1_ibmz15_2020} and Intel's QAT 4xxx~\cite{yuan2024intel} reduce latency and power through CPU integration; recent results on Xeon Platinum 8458P~\cite{wang2025reviving} highlight their superior small-chunk handling, though its scalability is limited. In-storage solutions like ScaleFlux CSD~\cite{scaleflux_csd2000} and Samsung SmartSSD~\cite{samsung_smartssd, tian2024scalable} minimize data movement through in-storage processing and achieve near-linear performance scaling by adding CSD drives. These architectures differ in cost efficiency, CPU offloading, performance, power, and deployment flexibility, illustrating that the optimal acceleration strategy must align with specific datacenter workloads and requirements. 

\subsection{Motivation} 
\textbf{In-storage CDPU design.} Although SSDs with integrated compression/decompression processing units (CDPUs) are emerging, there is a notable lack of research on their architecture and implementation. Sparse academic documentation of in-storage CDPU solutions has impeded both innovation and adoption. This gap is especially important since in-storage CDPUs can uniquely optimize data handling and efficiency through direct integration with SSD page management. However, key architectural trade-offs, performance impacts, and implementation challenges remain insufficiently explored, limiting progress in harnessing their full benefits.

\textbf{CDPU placement study with genuine ASIC devices.} Existing research is predominantly focused on peripheral and on-chip CDPU designs, such as Intel QAT in QZFS~\cite{qzfs2019} and IBM z15 NXU~\cite{b1_ibmz15_2020}, while in-storage CDPUs lack comprehensive real-world evaluation. The absence of empirical, hardware-based comparisons across peripheral, on-chip, and in-storage CDPUs leaves a critical gap in our understanding of optimal CDPU placement. Addressing this gap with systematic, hardware-based studies is essential to inform architectural choices and deployment strategies suited to diverse storage system requirements.

\begin{figure}
    \centering
    \includegraphics[width=\linewidth]{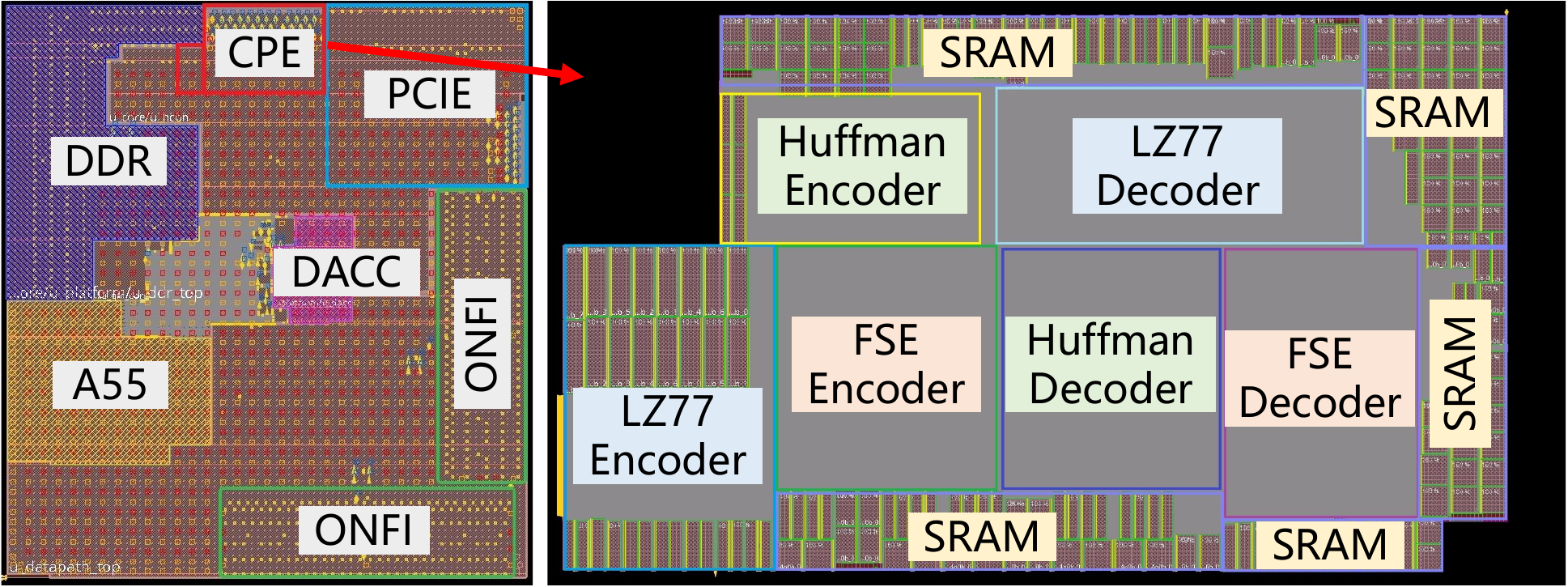}
    \vspace{-2em}

    \caption{SSD controller floorplan with DPZip (CPE) highlighted. DPZip block occupies 6mm$^2$ (4.5\%) of the 132mm$^2$ chip area.}
    \label{fig:dp800_floorplan}
\end{figure}

\section{Design and Implementation of DPZip}\label{design}
\subsection{DPZip Overview}
DPZip is a hardware-optimized ASIC implementation of dictionary and entropy encoders tailored for SSDs, operating directly on 4KB flash pages with LZ77~\cite{ziv1977universal} pattern matching and entropy coding using Huffman~\cite{huffman1952method} or FSE~\cite{Duda2013,collet2013fse}. To fit tight silicon budgets, DPZip uses a compact LZ77 hash table and limits Huffman codes to 11 bits. Integrated into a PCIe 5.0 SSD controller, DPZip processes (de)compression at 8 bytes per cycle, reaching up to 16GB/s throughput and $\sim$2$\mu$s 4KB transfer latency, while occupying about 4.5\% of the die (6 mm$^2$ of 132 mm$^2$). Efficient LZ77, dynamic Huffman, and FSE engines enable high bandwidth and low power, addressing challenges of SRAM overhead and performance fluctuations in (de)compression. Performance is further improved with two-stage matching and optimized short/long-range decompression. As shown in Figure~\ref{fig:dp800_floorplan}, DPZip is placed along the controller’s main interconnect near critical SoC modules and is organized into dedicated, SRAM-coupled LZ77, FSE, and Huffman units to minimize data movement and latency.

\subsection{Dictionary Encoder: LZ77}
\label{sec:lz77_encoder_decoder}

\subsubsection{\textbf{Basic Flow}}
The hardware-optimized LZ77 encoder operates in a pipelined sequence:

\begin{itemize}[leftmargin=*]
    \item \textit{\ul{Input Processing:}} The input stream is divided into blocks, each processed in groups of four consecutive positions. This grouping enables parallel processing, which is crucial for high throughput in ASIC implementations.
    
    \item \textit{\ul{Hashing and Lookup:}} For each position, a hardware-friendly hash is computed to index a fixed-size, multi-slot hash table holding multiple candidate locations. This two-level scheme first provides a fast, coarse candidate selection.
    
    \item \textit{\ul{Match Retrieval and History Match:}} Retrieved candidates undergo a byte-by-byte \emph{history match} to determine exact match length. The combination of rapid hash lookup and fine-grained comparison minimizes comparisons and pipeline stalls.
    
    \item \textit{\ul{Sequence Token Generation:}} When a valid match surpasses the minimum threshold, the encoder outputs a token (literal length, match length, offset) and advances the input pointer by the match length. Meanwhile, the hash table is updated in parallel, ensuring seamless processing for subsequent positions.
\end{itemize}

\subsubsection{\textbf{Challenges}}
High-performance LZ77 ASIC compression requires carefully balancing compression ratio, throughput, latency, and silicon area, with on-chip SRAM as the critical constraint due to its area and power implications. Achieving area efficiency necessitates highly compact hash tables, while maximizing throughput depends on minimizing hash-table access and wiring delays within strict area and power budgets. Pipeline stalls from long matches are mitigated by replicating match units, but deeper pipelines introduce control complexity and may increase latency due to pipeline bubbles. In SSD controllers, these challenges intensify as compression must share limited die area with flash controller and ECC modules, resulting in truncated hash chains to fit SRAM limits. Software-oriented algorithms like Zstd~\cite{zstd} and Deflate~\cite{deutsch1996rfc1951}, which achieve good compression ratios using large sliding windows (32--128KB) and complex parsing, rely on pointer-heavy operations that are inefficient for hardware. Recent FPGA/ASIC solutions~\cite{b1_ibmz15_2020,beezip2024asplos} boost throughput via match unit replication and pipeline depth at the expense of increased LUT and SRAM usage. Consequently, hardware algorithms must aggressively optimize both computational overhead and memory use, making efficient design extremely challenging.

\subsubsection{\textbf{Encoding Designs}}\label{lz77_encode}
The LZ77 restricts hash table size, splits matching into two stages (fast hash check, then full compare), performs  lazy matching in an efficient but hardware-friendly manner, and relies on pipeline concurrency and partial prefetch to sustain throughput.

\textit{\ul{SRAM-Optimized Hash Table.}}
The LZ77 encoder maintains a small, bounded hash table array, in which each entry holds only a few candidate positions. During each iteration, the code computes $\mathit{Hash0}$ and $\mathit{Hash1}$ for the current 4-byte block, looks up those indices in the table, and compares them against stored positions to locate potential repeats. If a match is confirmed, it is represented by a $\langle LL, ML, \mathit{Off} \rangle$ tuple, where $LL$ represents the literal length, $ML$ denotes the match length, and $\mathit{Off}$ indicates the distance to the reference data. If not, the data is treated as literal bytes.

To limit the number of stored positions per hash index, entries are stored in a circular FIFO manner. Consequently, older entries are naturally evicted without complicated data structure management.
This design ensures the hash table fits in a minimal on-chip SRAM, reducing hardware overhead while preserving effective detection of recent matches.

\textit{\ul{Two-Level Match-Processing.}}
DPZip processes each 4-byte word by computing two 1-byte hash values to quickly index stored references. If the hash indicates a possible match, a byte-wise verification confirms the match length, preventing false positives that could disrupt the pipeline. The bounded hash table is then updated, removing old entries and inserting the latest one, either per iteration or every 4 bytes. This incremental update maintains an up-to-date match dictionary while reducing hardware complexity.

\textit{\ul{Lazy Matching.}} 
DPZip's LZ77 employs partial-lazy matching by skipping ahead 4 bytes when no match is found. Upon finding a potentially suboptimal match, the pipeline accepts it without backtracking, enabling a first-fit match policy. This approach slightly harms compression ratio but significantly streamlines logic and sustains high pipeline throughput.

\subsubsection{\textbf{Decoding Designs}}
The ASIC implementation of the LZ77 decoder processes two fundamental data types: literal bytes and sequences defined by $\langle LL, ML, \mathit{Off} \rangle$ tuples. Maintaining consistent decoding performance poses a significant challenge.

\textit{\ul{Overlap-Resilient Design for Short Offsets}}. The primary technical challenge lies in efficiently managing memory access patterns, particularly for overlapping match regions where the read and write addresses closely coincide. To address this, the design employs a dual-buffer architecture with separate literal and history buffers, implemented using dual-port SRAM. 
To further mitigate the access latency, it adopts a small register-backed recent-data buffer (typically 256 bytes) to enable immediate data access for short-offset matches without SRAM read latency.

\textit{\ul{Two-Level Copy-Processing}}. The architecture features a dual-pipeline system controlled by a finite state machine (FSM). The literal pipeline handles direct byte transfers from the literal buffer, while the match pipeline manages data replication from the history buffer. This separation enables parallel processing and optimizes throughput. To ensure consistent bandwidth, the design uses prefetching for longer offsets and bypass logic for recent writes.

\subsection{\textbf{Entropy Encoders: Huffman and FSE}}
DPZip utilizes canonical Huffman coding to minimize memory usage by representing codebooks with compact code lengths instead of pointer-based trees. Sorting symbols by code length and value enables the decoder to efficiently reconstruct codes, facilitating direct table look-up and eliminating the need for tree traversal. This approach reduces on-chip memory and wiring complexity, providing a streamlined, area-efficient solution well suited for ASIC-based compression hardware.

\ul{\textit{Efficient Huffman Tree Canonization.}}
DPZip reshapes the Huffman tree into a hardware‑bounded, pipeline‑neutral procedure that canonicalises any over‑deep Huffman tree while enforcing a ceiling (e.g., 11‑bit). Unlike the software implementation with variable‑length ``cost‑repayment'' loop~\cite{facebook_zstd}, data‑dependent branching, and mutable \texttt{rankLast[]} table that make timing unpredictable, our ASIC-oriented design folds the algorithm into three latency-stable stages.

\begin{enumerate}[label=\arabic*.\,,leftmargin=*]
  \item \textbf{Leaf Scan \& Cap.}  A single forward pass streams all 256 symbols. Leaves deeper than~11 are clipped on‑the‑fly, while counters tally the total leaves ($N$) and the resulting hole deficit ($k$). Decisions depend only on the current symbol's \texttt{nbBits}, so the stage is purely combinational and stall‑free.

  \item \textbf{Deterministic Redistribution.}  A compact FSM walks through levels~10$\,\rightarrow\,$1, promoting just enough leaves each cycle to absorb $k$. Arithmetic is limited to shifts and increments, shortening the critical path by \(\sim\!23\,\%\) and eliminating wide multipliers/dividers.

  \item \textbf{Logarithmic Hole Repair.}  Any residual deficit is propagated upward, halving each cycle. For a 256‑symbol alphabet the loop terminates in at most $\lceil \log_{2} k \rceil \le 8$ iterations.
\end{enumerate}

Each stage is timing‑balanced, enabling closure at \textbf{1GHz} in a 12~nm process. The worst‑case schedule is tightly bounded:
\[
  T_{\max} = 256 (\text{scan}) + 10 (\text{redistrib.}) + 8(\text{repair}) = 274\,\text{cycles}
\]

The FSE hardware encoder/decoder is fully compatible with the software implementation~\cite{collet2013fse} in Zstd. The integrated circuit leverages specialized finite state machines and deeply pipelined datapaths to deliver deterministic latency and high throughput.

\begin{figure}[t]
    \centering
    \includegraphics[width=7cm]{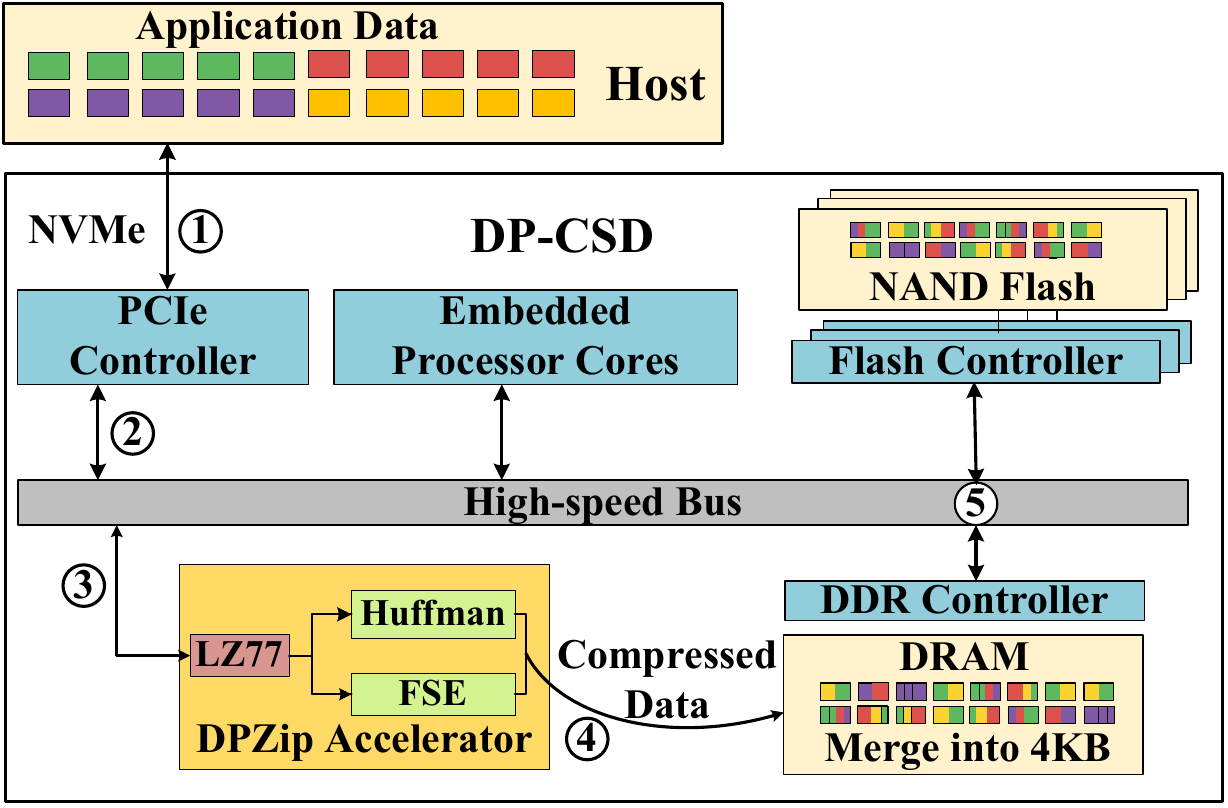}
    \vspace{-0.5em}
    \caption{DPZip accelerator and its integration in DP-CSD. }
    \label{fig:DPZip_integrate}
\end{figure}

\section{DP-CSD: DPZip-Powered SSD}\label{dpcsd}

\subsection{SoC Microarchitecture and Interconnection}
Figure~\ref{fig:DPZip_integrate} demonstrates the SSD controller architecture, emphasizing how the on‐chip DPZip accelerator is integrated.

\textbf{Processing units.}
Several specialized processing blocks drive efficient execution in the design. Embedded ARM cores execute firmware for critical tasks, including Flash Translation Layer (FTL), garbage collection, and wear-leveling. The Queue Manager (QM) retrieves NVMe commands from host queues and directs data flow. DPZip integrates hardware-accelerated compression and decompression in the data path, reducing NAND write amplification and increasing effective storage capacity.

\textbf{Control and data flows.}
SSD data and control flows are tightly orchestrated for optimal performance. When the host issues a read or write command via PCIe, the QM retrieves and processes the NVMe command. For writes, incoming host data is sent into on-chip Shared Buffer Memory (SBM) in high-speed SRAM through DMA, where it is temporarily staged before DPZip performs compression. The compressed output is then transferred to the flash controller (FLC) and programmed into NAND flash. For reads, the FLC fetches data from NAND and places it in the SBM; DPZip then decompresses the data, which is finally sent back to the host.

\subsection{CDPU-Compatible SSD FTL Firmware}
\begin{figure}[t]
    \centering
    \vspace{-0.5em}
    \includegraphics[width=0.9\linewidth, trim=1cm 4.8cm 1cm 4cm, clip]{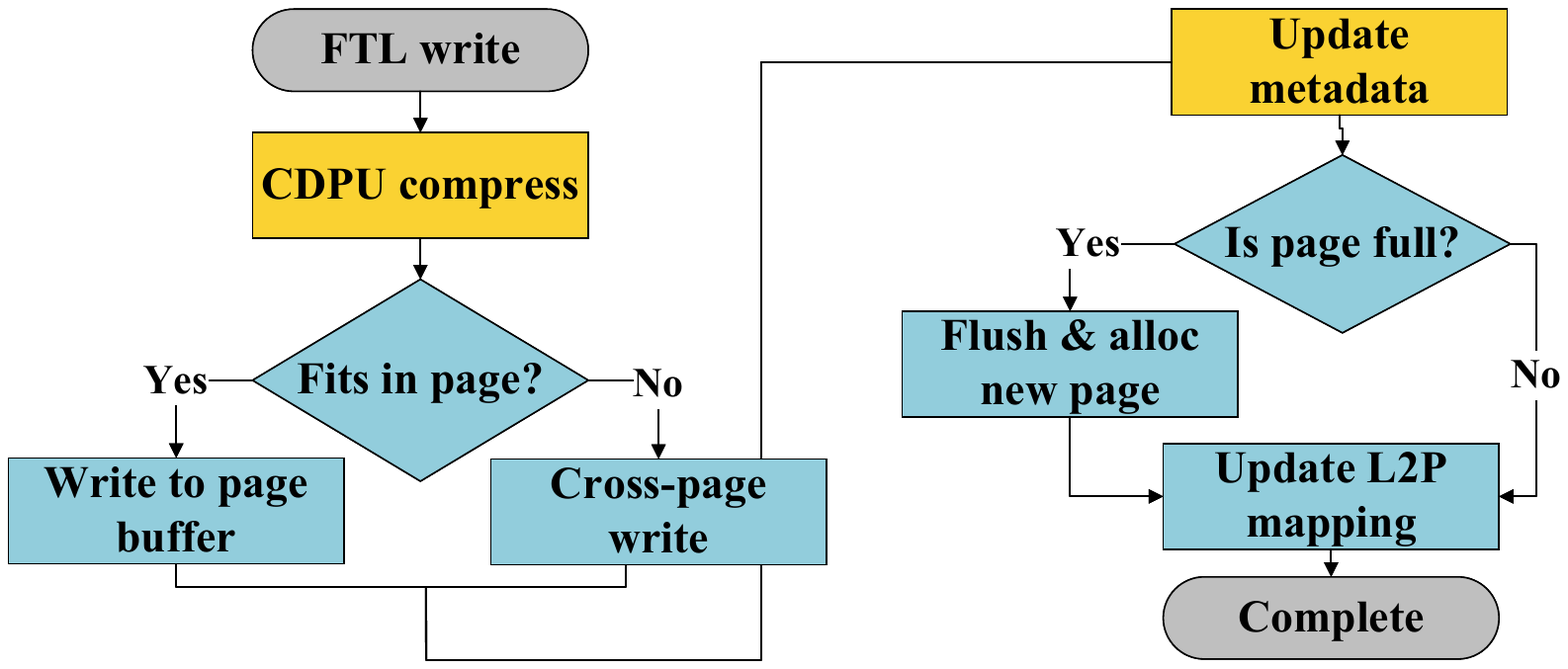}
    \vspace{-0.5em}
    \caption{Write flow of DP-CSD with DPZip integration.}
    \label{fig:dpcsd_write_flow}
\end{figure}

Integrating data compression inside SSDs introduces significant challenges. We adapt the FTL firmware for IO routing, metadata, and data management. As illustrated in Figure~\ref{fig:dpcsd_write_flow}, DP‑CSD embeds the DPZip engine within a page-aware control flow inside the FTL write path. On host writes, data is compressed at line rate before flash write. If the compressed data fits within a single page, it is buffered directly; if not, it is split across pages with sequential mapping to prevent fragmentation. Metadata is updated accordingly: full pages are committed and trigger new allocations, while partial pages only advance their write pointers. The page mapping table is atomically updated, keeping metadata operations off the performance-critical path and reducing write amplification.

The FTL employs a log-structured architecture that tightly couples compression with address mapping to optimize storage utilization. Host data undergoes inline compression, with compressible segments packed into flash pages to minimize storage footprint while incompressible data is stored uncompressed to avoid management overhead. When compressed segments exceed the remaining capacity of the current page, they are flagged and continued on the subsequent page. The system buffers associated metadata, applies error correction coding (BCH or LDPC), and commits it to flash alongside multi-page parity protection and supercapacitor-backed data consistency mechanisms.

Compression increases the effective storage capacity, which is the actual volume of user data an SSD can accommodate after compression, enabling storage beyond the device's raw physical limits for compressible workloads (e.g., doubling capacity with a 50\% compression ratio). DP-CSD allows users to configure and expose the SSD's effective capacity to the host operating system, calibrated to expected compression ratios, thereby optimizing space utilization.

The FTL maintains mapping coherence by updating its in-DRAM logical-to-physical (L2P) page mapping table with address translations and invalidating obsolete locations for garbage collection. Logical pages may span multiple physical pages due to compression boundaries, introducing read penalties. The DPZip engine performs inline decompression during read operations to maintain transparent data access.

Note that DPZip is inherently protocol-agnostic and does not rely on the NVMe specification. Our current DP-CSD implementation targets enterprise NVMe environments, leveraging NVMe's prevailing status in high-performance storage. Adapting DPZip for consumer SSDs with SATA interfaces would necessitate reengineering the SSD controller and reducing the parallelism of DPZip modules to match SATA's sub-1GB/s bandwidth, in contrast to its native throughput exceeding 10GB/s. Such modifications are essential to prevent hardware overprovisioning and ensure cost-effectiveness in consumer-grade devices.

\section{Evaluation}\label{evaluation}
We benchmark DPZip (\textbf{in-storage} CDPU) against established compression solutions across CPU-based software implementations (Snappy~\cite{snappy}, Zstd~\cite{zstd}, Deflate~\cite{deutsch1996rfc1951}) and Intel QAT-based \textbf{peripheral} and \textbf{on-chip} CDPUs~\cite{website:qat,qzfs2019,yuan2024intel}. 
The evaluation encompasses throughput, latency, compression ratio, power consumption, energy efficiency, and scalability under single-threaded and multi-threaded workloads.

\subsection{Evaluation Methodology}

\begin{figure}[t]
    \centering
    \includegraphics[width=8.5cm]{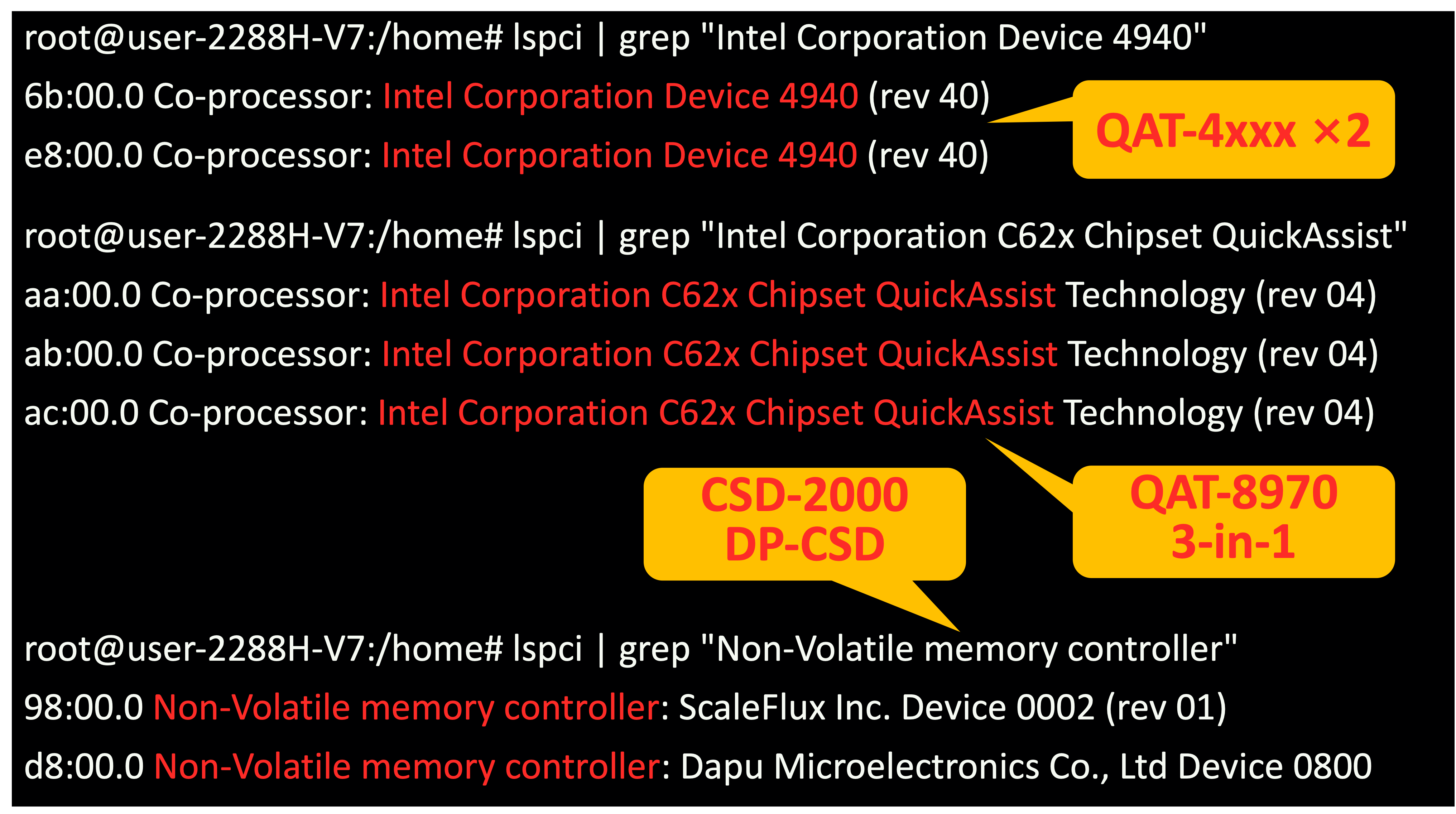}
    \vspace{-1em}
    \caption{The evaluation testbed features four hardware compression devices: dual embedded \textit{QAT 4xxx} engines (one per CPU socket), a \textit{QAT 8970} PCIe card with an ensemble of three co-processors, \textit{CSD 2000} with an FPGA compression engine, and \textit{DP-CSD} implementing DPZip.}
    \label{fig:devices}
    \vspace{-1em}
\end{figure}

Our testbed, illustrated in Figure~\ref{fig:devices} and detailed in Table~\ref{table:testbed}, comprises an xFusion 2288H V7 dual-socket server running Ubuntu 22.04.4 with Intel Xeon Platinum 8458P CPUs (44 cores/88 threads at 2.7 GHz, 82.5 MB cache) and 256 GB DDR5-4800 ECC memory. The system integrates four distinct hardware compression architectures: two embedded Intel QAT-4xxx accelerators (one per CPU socket), an Intel QAT 8970 PCIe card enumerating as three co-processors (PCIe 3.0 $\times$16), a ScaleFlux \textit{CSD 2000} with FPGA compression engine (PCIe 3.0 $\times$4), and a DapuStor DP-CSD implementing DPZip (PCIe 5.0 $\times$4). As demonstrated by device enumeration (Figure~\ref{fig:devices}), all CDPUs are deployed on the same platform.

We select Intel Xeon CPUs as the sole commercially available on-chip CDPU solution accessible to our research, with IBM mainframe processors~\cite{b1_ibmz15_2020} being the only alternative but unavailable for evaluation. The Intel QAT 8970 serves as our peripheral CDPU representative due to its architectural similarity to other PCIe accelerators and SmartNIC DPUs, while also enabling generational comparison as the predecessor to the on-chip QAT 4xxx. For in-storage CDPUs, we employ both the pioneering CSD 2000 FPGA solution (PCIe 3.0) and the advanced DP-CSD ASIC implementation (PCIe 5.0) to establish generational performance trajectories across both FPGA and ASIC architectures.

\begin{table}[t]
    \caption{Testbed configuration, evaluation targets, and metrics: “Local/Remote” refer to intra-/inter-NUMA memory performance; “C/D” denotes compression/decompression operations; “Interconnect” specifies data transfer path to CDPU during compression.}
    \vspace{-1em}
    \centering
    \scriptsize
    \setlength{\tabcolsep}{0pt} 
    \renewcommand{\arraystretch}{1.1} 
    \begin{tabular}{@{}llccc@{}}
        \toprule
        & & \multicolumn{2}{c}{\textbf{Local/Remote DRAM}} & \textbf{CPU/Cache Specification} \\
        \textbf{Server} & \#ch DDR & Lat (ns) & BW (GB/s) & \#Cores, L1D/L2/L3 \\
        \midrule
        SPR2S & 4$\times$DDR5 & 110/198 & 128/108 & 88 (2.7GHz), 80KB-2MB-80MB \\
        \midrule
        \midrule
        \textbf{CDPU} & \textbf{\#Instance} & \textbf{Placement} & \textbf{Interconnect} & \textbf{Algorithm/Specification} \\
        \midrule
        QAT 8970 & 3-in-1 ASIC & Peripheral & PCIe 3.0$\times$16 & Deflate, 66/160Gbps (C/D)~\cite{website:qat} \\
        QAT 4xxx & 2$\times$ ASIC & CPU on-chip & CMI & Deflate, 160/160Gbps (C/D)~\cite{yuan2024intel} \\
        CSD 2000 & 1$\times$ FPGA & In-storage & FPGA AXI & Gzip, 20/24Gbps (C/D)~\cite{scaleflux_csd2000} \\
        DPZip & 1$\times$ ASIC & In-storage & Chiplet AXI & Zstd variant, 128/160Gbps (C/D) \\
        \midrule
        \midrule
        \textbf{Target} & \multicolumn{3}{c}{\textbf{Evaluation Metric}} & \textbf{Benchmark Tool} \\
        \midrule
        CDPU & \multicolumn{3}{c}{Throughput/Latency/Compression Ratio} & QATzip~\cite{intel_qatzip}/FIO~\cite{fio} \\
        Application & \multicolumn{3}{c}{Throughput/Latency/Execution Time} & RocksDB/Btrfs/ZFS \\
        System & \multicolumn{3}{c}{CPU Utilization/Power Efficiency} & RocksDB/Btrfs \\
        \bottomrule
    \end{tabular}
    \label{table:testbed}
\end{table}

Evaluation proceeds across three levels. At the device level, microbenchmarks measure compression and decompression throughput, latency and power consumption. At the application level, end-to-end performance is gathered from RocksDB and Btrfs. At the system level, we record CPU and memory utilization along with total power draw to assess overall efficiency. Compression ratios are obtained with Deflate, Zstd, LZ4 and Snappy on the Silesia dataset~\cite{silesia_corpus}, which has been widely used in previous work~\cite{b1_ibmz15_2020, chen2024hacsd, karandikar2023cdpu, HyperCompressBench, beezip2024asplos}, to reflect realistic data patterns. Deflate and Zstd are both executed at level 1 to align with DPZip.

\subsection{CDPU Capabilities}
In this section, we present a comprehensive analysis of multiple compression schemes, quantifying their performance across four critical metrics: throughput, latency, compression ratio, and system power efficiency. The primary evaluation focuses on a 4KB compression granularity, selected specifically to maintain alignment with standard SSD page sizes. To provide broader analytical context, we extend our investigation to include 64KB data blocks, allowing for a direct comparison of performance characteristics across different granularities. Since QAT accelerators implement the Deflate compression algorithm, we also employ Deflate as the CPU software compression baseline, ensuring a fair comparison across the board. All testing parameters remain consistent across the evaluated compression solutions, including CPU-based approaches and three ASIC-based solutions: QAT 8970, QAT 4xxx, and DPZip.

\begin{figure}[t] \centering 
    \begin{subfigure}[b]{0.47\linewidth} \centering \includegraphics[height=3.5cm]{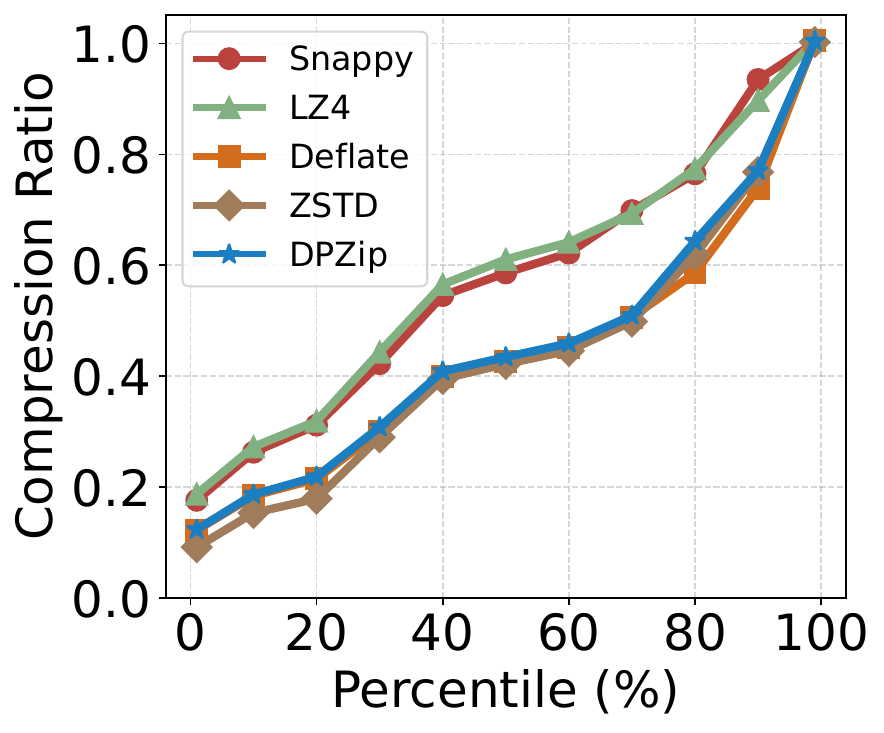} 
    \vspace{-1.5em} \caption{Compression in 4KB.} \label{fig:DPZip-vs-Zstd_4kb} 
    \end{subfigure} 
    \hfill 
    \begin{subfigure}[b]{0.47\linewidth} \centering \includegraphics[height=3.5cm]{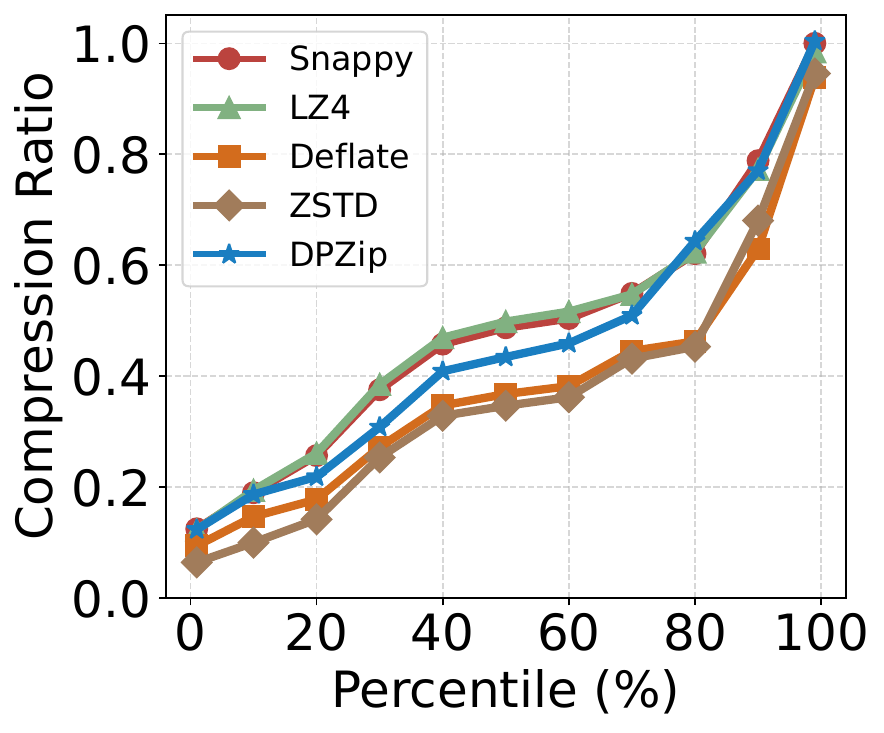} 
    \vspace{-1.5em} \caption{Compression in 64KB.} \label{fig:DPZip-vs-Zstd_64kb} 
    \end{subfigure} 
    \vspace{-1em} 
    \caption{Compression ratio distributions of the Silesia dataset are compared across various compression algorithms.} 
    \label{fig:DPZip_cr}
\end{figure}

\subsubsection{\textbf{Compression Ratio}}
Modern storage architectures feature hierarchical granularities that critically impact IO performance. Legacy HDDs use 512B logical sectors, while contemporary SSDs internally organize data in 4KB pages, creating granularity mismatches that trigger costly read-modify-write operations for 512B requests. Modern operating systems mitigate this by issuing 4KB-aligned requests matching SSD page organization. Additionally, high-capacity SSDs adopt larger indirection units (16–64KB) to reduce DRAM overhead in flash translation layer mapping tables~\cite{zhou2024csal}.

DP-CSD addresses this evolving landscape through a dual-granularity architecture that ensures broad compatibility. The interface layer supports variable logical block sizes from 512B to 64KB per NVMe namespace specification~\cite{nvme2025}, while the compression layer maintains fixed 4KB compression granularity independent of logical block or indirection unit sizes. This design decouples compression management from storage device heterogeneity.

Figure~\ref{fig:DPZip_cr} shows compression ratios (lower is better) for various algorithms at both 4KB and 64KB granularity, corresponding to typical logical block size of SSDs and the larger indirection units adopted by high-capacity SSDs.

\textbf{Finding 1. SSD page compression:}
\textit{At 4KB granularity (conventional SSD page size), DPZip achieves a compression ratio comparable to Deflate and Zstd, and significantly surpasses lightweight compressors such as Snappy and LZ4.}
On the Silesia dataset~\cite{silesia_corpus}, Deflate and QAT 8970 yield 43.1\% and QAT 4xxx achieves 42.1\%. DPZip attains 45\%, slightly worse due to its LZ77 design that favors resource efficiency with minimal penalty in compression ratio (\S\ref{lz77_encode}). Compression ratio is sensitive to chunk size, with all algorithms showing reduced efficacy at 4KB granularity. Among traditional compressors, Deflate and Zstd deliver the best results, but DPZip closely tracks their performance across all percentiles. At 64KB, QAT's compression ratio improves to 36–38\%, while DPZip, processing all requests as 4KB pages, maintains a stable ratio independent of IO size. Across both granularities, even without optimization for large pages, DPZip consistently outperforms Snappy and LZ4 lightweight algorithms in compression ratios.

\begin{figure}[t]
    \centering
    \begin{minipage}[t]{0.48\textwidth}
        \centering
        \includegraphics[height=3.5cm]{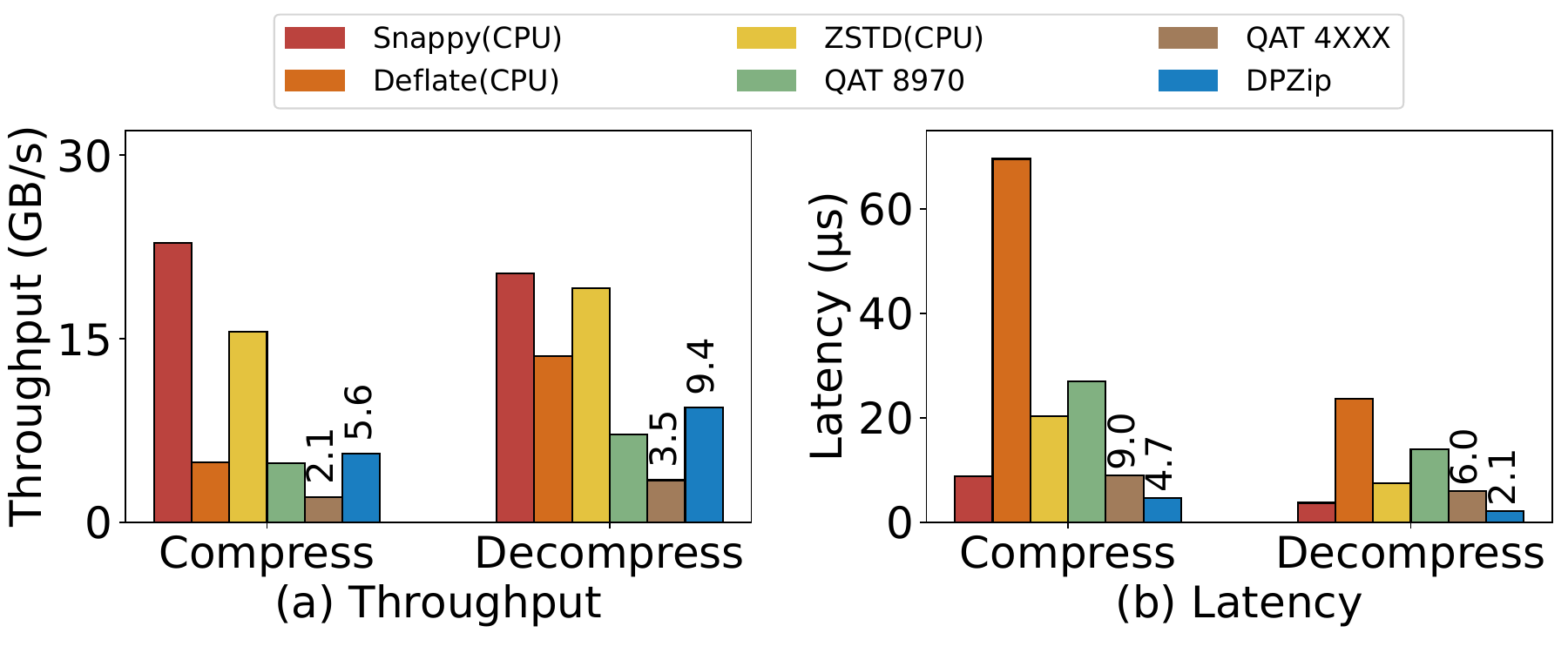}
        \vspace{-1em}
        \caption{Compression at 4KB granularity. \textit{CSD 2000} is excluded since its FPGA CDPU metrics are not reported. 
        }
        \label{fig:microbenchmark-4kb}
    \end{minipage}%
    \hfill
    \vspace{1em}
    \begin{minipage}[t]{0.48\textwidth}
        \centering
        \includegraphics[height=3.5cm]{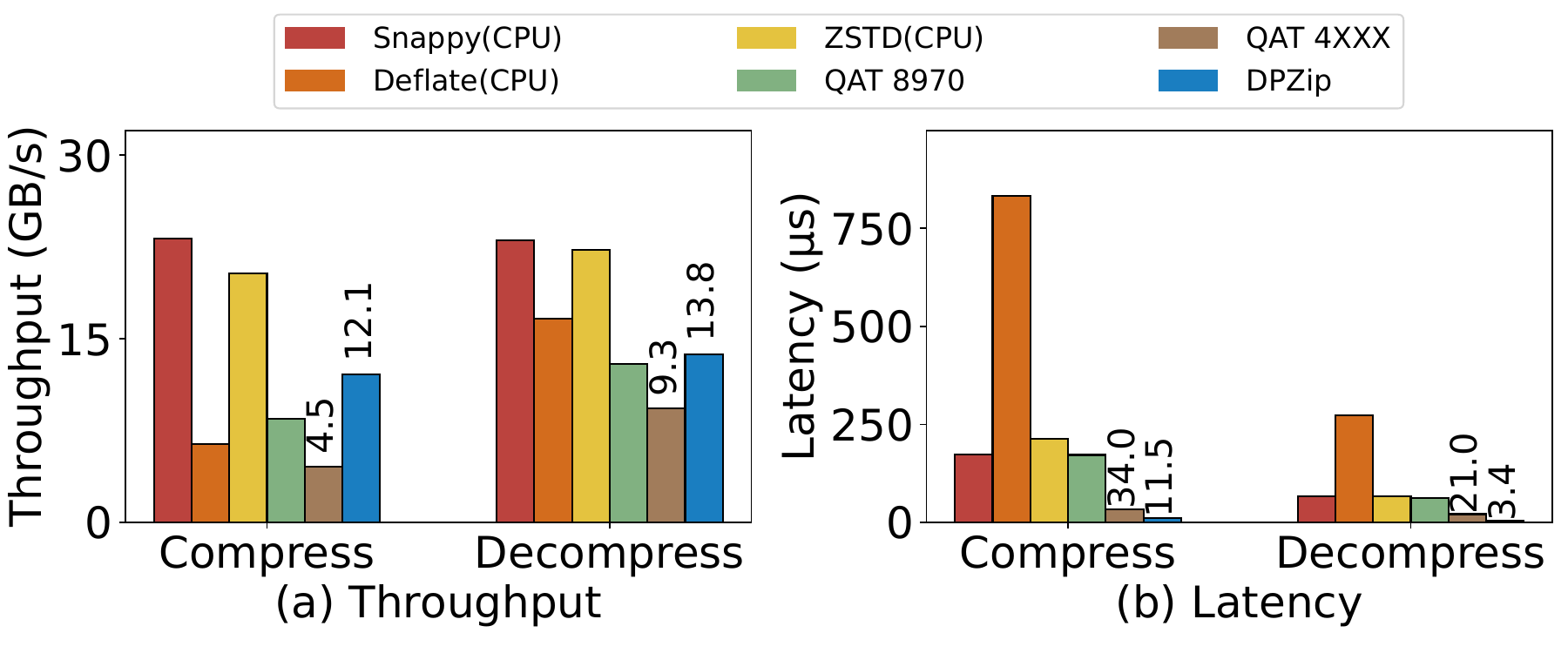}
        \vspace{-1em}
        \caption{Compression at 64KB granularity.}
        \label{fig:microbenchmark-64kb}
    \end{minipage}
\end{figure}

\subsubsection{\textbf{Throughput}.}
Figure~\ref{fig:microbenchmark-4kb}(a) summarizes throughput for 4KB data blocks. With 88 CPU threads, Deflate software achieves 4.9GB/s compression and 13.6GB/s decompression, while a PCIe QAT 8970 accelerator offers 5.1GB/s and 7.6GB/s, respectively. The latest QAT 4xxx on-chip accelerator, despite its integration, delivers only 4.3GB/s compression and 7GB/s decompression—trailing the legacy QAT 8970. The DPZip in-storage accelerator leads among ASICs with 5.6GB/s compression and 9.4GB/s decompression.

\textbf{Finding 2. IO granularity impact:} \textit{Larger IO granularities (e.g., 64KB as the QAT hardware buffer size) significantly boost CDPU throughput.} As shown in Figure~\ref{fig:microbenchmark-64kb}(a), processing 64KB chunks with the Deflate software algorithm increases throughput by 30\% compared to 4KB chunks. Hardware CDPUs achieve even higher performance gains than software, with compression throughput increased by 74–120\% and decompression by up to 177\%. Utilizing three parallel DP-CSDs with 64KB chunks further boosts aggregate compression throughput to 37.5GB/s. Larger requests reduce context switches and queuing events, improving PCIe bandwidth utilization, but larger compression granularities can hurt read/update performance since more data must be decompressed for partial accesses.  

The CPU's general-purpose architecture enables flexible algorithm switching, allowing high throughput with lightweight compressors such as Snappy (22.8GB/s compression, 20.3GB/s decompression). In contrast, ASIC accelerators typically lack algorithm-switching capability. However, ASICs like QAT 8970 and DPZip support linear scalability through parallel deployment; for example, three DPZip-enabled computational storage drives achieve 16.3GB/s compression and 20.9GB/s decompression at 4KB granularity, exceeding even highly optimized CPU implementations. Overall throughput depends not only on algorithm and accelerator efficiency but also on interconnect bandwidth and data transfer mechanisms, such as DMA and queuing.

\subsubsection{\textbf{Latency}}
Compression latency is a critical factor in storage systems, as it can directly limit IO performance. As Figure~\ref{fig:microbenchmark-4kb}(b) shows, CPU-based software Deflate incurs latencies up to 70$\mu$s for 4KB data blocks, substantially higher than the typical SSD write latency of sub-10$\mu$s (internal buffered write). This substantial gap forces databases to favor lower-latency, but less space-efficient algorithms like LZ4 and Snappy over high-compression ratio counterparts.

\textbf{Finding 3. Memory proximity benefit:} \textit{On-chip CDPUs benefit from memory proximity (e.g., CMI, DDIO/LLC) and deliver much lower latency than PCIe-attached accelerators.} In general, hardware accelerators significantly reduce latency, with QAT 8970 achieving 28$\mu$s compression and 14$\mu$s decompression latencies. QAT 4xxx, leveraging superior memory access performance, further reduces these to 9$\mu$s and 6$\mu$s.

\begin{figure}[t]
    \centering
    \includegraphics[width=8.5cm]{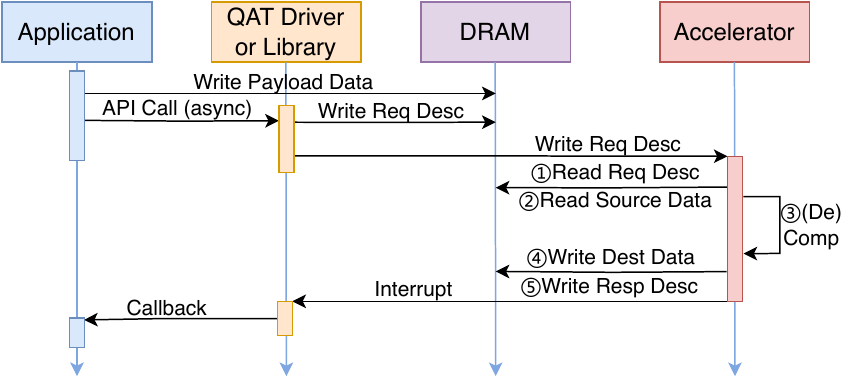}
    \vspace{-2em}
    \caption{Request processing flow for both QAT 8970 and QAT 4xxx architectures.}
    \label{fig:timeline}
\end{figure}

Figure~\ref{fig:timeline} presents the QAT processing flow. The accelerator pipeline initiates with an asynchronous API call triggering the driver to generate and enqueue request descriptors. The accelerator DMA-reads these descriptors from host memory, potentially hitting the last-level cache (LLC) if data direct I/O (DDIO)~\cite{Intel-DDIO} is available for the on-chip CDPU. DDIO enables chiplets to directly access the CPU's LLC, bypassing main memory to reduce latency. The accelerator then DMA-reads the source payload, again leveraging DDIO to bypass DRAM when data resides in the LLC. The (de)compression pipeline executes entirely on-accelerator, incurring only compute latency. Upon completion, the accelerator DMA-writes results to the destination buffer, with DDIO ensuring direct LLC placement for immediate CPU visibility, before delivering interrupts for driver interrupt service routine (ISR) handling and application callbacks.

\begin{figure}[t]
    \centering
    \includegraphics[width=8.5cm]{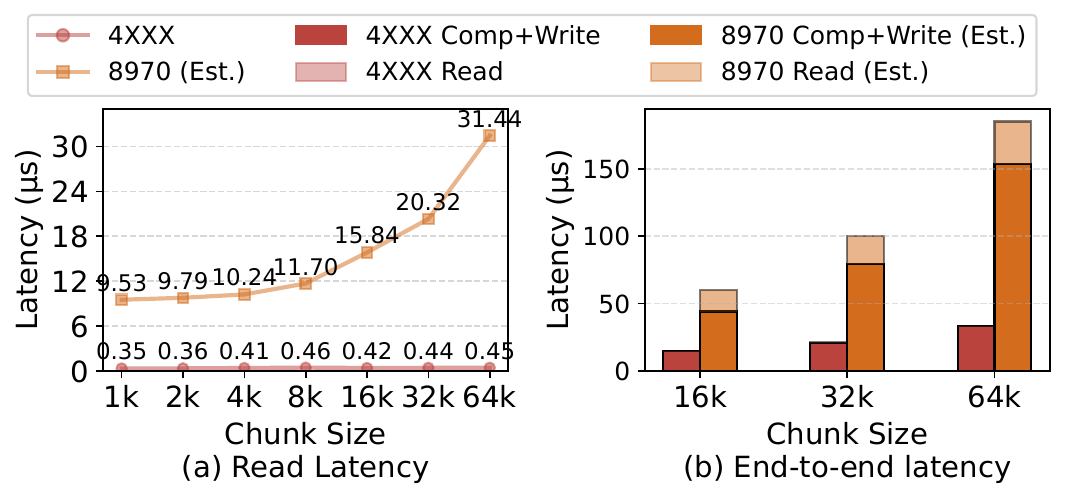}
    \vspace{-2em} 
    \caption{Latency analysis under various chunk sizes.} 
    \label{fig:combined}
\end{figure}

Figure~\ref{fig:combined} reveals critical performance differences through QAT processing latency breakdown. QAT 4xxx telemetry demonstrates exceptionally low read latencies (448ns for 64KB), enabled by DDIO. Since QAT 8970 lacks IO telemetry support, we estimate its DMA latency using PCIe SSD controller memory buffer (CMB) experiments, as both utilize PCIe DMA for host data exchange. By varying host access granularity to the CMB, we determine that QAT 8970's PCIe DMA latency substantially exceeds QAT 4xxx (up to 70$\times$ higher), as shown in Figure\ref{fig:combined}(a).

Despite QAT 8970's superior throughput (Figures~\ref{fig:microbenchmark-4kb} and \ref{fig:microbenchmark-64kb}) via hardware parallelism (three co-processors per device), Figure~\ref{fig:combined}(b) demonstrates its end-to-end processing latency remains 3–5$\times$ higher than QAT 4xxx. These results underscore how accelerator placement and efficient IO data paths critically impact latency reduction and storage accelerator performance optimization.

\textbf{Finding 4. Benefits of in-storage integration:} \textit{Tightly coupled in-storage CDPU (e.g., DPZip) minimizes latency and maximizes throughput by eliminating costly host-CDPU data movement.}
DPZip performs compression directly along the write path, thereby eliminating relatively slow memory copy operations. This design enables DPZip to achieve exceptionally low latencies, 4.7$\mu$s for compression and 2.6$\mu$s for decompression. In comparison, CPU-optimized software solutions such as Zstd and Snappy exhibit higher latencies, with Zstd requiring 20.4$\mu$s and 7.4$\mu$s for compression and decompression, respectively, and Snappy requiring 8.9$\mu$s and 3.8$\mu$s. Consequently, DPZip not only provides high compression ratios but also delivers latencies that are approximately 50\% lower than the fastest software-based alternatives. These results underscore the substantial performance benefits achievable through tightly coupled, in-storage CDPU architectures.

\begin{figure}[t]
    \centering
    \vspace{-1em}
    \includegraphics[width=8.5cm]{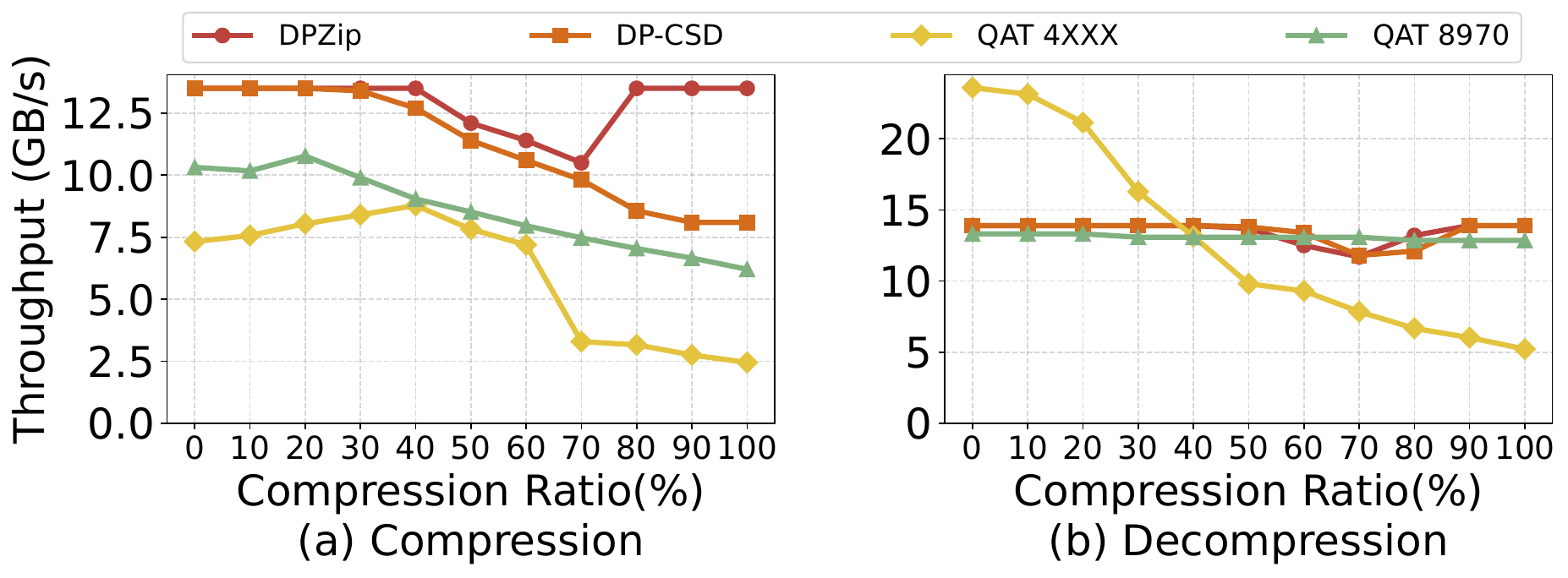}
    \vspace{-2em}
    \caption{Throughput under various compression ratios.}
    \label{fig:pattern}
\end{figure}

\subsubsection{\textbf{Data Pattern and CDPU Behaviors}}
CDPUs compress data by detecting and encoding redundant patterns, but their effectiveness declines on incompressible inputs, where unsuccessful pattern matching raises computational overhead and degrades throughput. 

\textbf{Finding 5. DPZip performance robustness:} \textit{DPZip exhibits robust throughput, with only minor performance declines (within 15\%) across the full range of compressibility patterns, while QAT 4xxx suffers pronounced throughput drops on incompressible data.} Hardware CDPUs typically include a decompression verification step after compression, causing decompression slowdowns to propagate directly to end-to-end throughput. This interdependence manifests in both DPZip and QAT 4xxx, with compression performance degrading markedly when decompression bottlenecks occur. As shown in Figure~\ref{fig:pattern}, the QAT 4xxx experiences a degradation of 67\% in compression throughput and 77\% in decompression throughput when processing incompressible data. In contrast, DPZip maintains throughput variation within 15\% across the same conditions. This result underscores the advantage of DPZip LZ77's ability to avoid unrewarded and computationally expensive matching attempts. Furthermore, we observe that when decompression throughput recovers, specifically in the range of 80--100\% compression ratio, compression throughput increases correspondingly. This correlation highlights the influence of decompression verification processes on overall compression performance. Notably, QAT 4xxx exhibits much steeper degradation than QAT 8970 as data compressibility drops, revealing increasing challenges for on-chip CDPU scaling.

Data compression generates variable-length data, requiring the SSD FTL to perform operations such as concatenation, padding, and page splitting to optimize storage utilization. These operations incur overhead, and when compressed data spans multiple physical pages, read amplification occurs because a single logical read may access multiple physical pages. To analyze how storage medium and data layout affect compression performance, we compare DP-CSD (a complete system using NAND flash) with DPZip (identical execution path but substituting DRAM for NAND storage). As demonstrated in Figure~\ref{fig:pattern}, DP-CSD suffers more acute performance degradation than DPZip on poorly compressible data. Notably, while DPZip demonstrates performance recovery at 80--100\% compression ratios due to hardware algorithm characteristics, DP-CSD shows no such rebound. These results emphasize that storage medium and data layout critically influence end-to-end CSD performance.

\subsection{Performance of Applications Powered by CDPUs}
Microbenchmarks highlight the impact of accelerator placement on compression and system metrics but fall short in two areas. First, accelerator invocation methods vary: for example, DPZip lacks a standalone interface and must be measured using the FIO benchmark, introducing IO stack overheads that distort true compression costs. Second, the adaptation required at the application and system layers can mask the accelerators' raw performance benefits, making end-to-end evaluation in real applications essential.

\begin{figure}[t]
    \centering
    \includegraphics[width=8.3cm]{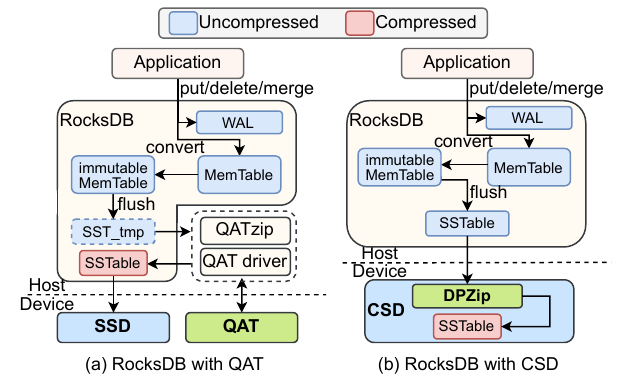}
    \vspace{-1.2em}
    \caption{QAT integration in RocksDB SSTable write path. DPZip/DP-CSD is application-transparent.}
    \label{fig:rocksdb}
\end{figure}

To ensure direct comparability, we focus on RocksDB~\cite{rocksdb_github} and Btrfs~\cite{rodeh2013btrfs}, two publicly available applications supporting both QAT 8970 and QAT 4xxx accelerators. DP-CSD, with integrated DPZip, functions as a standard NVMe SSD that is application-transparent and supports any database or filesystem without requiring modification. As shown in Figure~\ref{fig:rocksdb}, QAT accelerators are tightly integrated into application write paths, SSTable writes in RocksDB and pre-flush compression in Btrfs, requiring deep coordination with each system. In contrast, DPZip operates transparently, eliminating the need for such application or filesystem adaptation.

\begin{figure}[t]
    \centering
    \includegraphics[width=8.5cm]{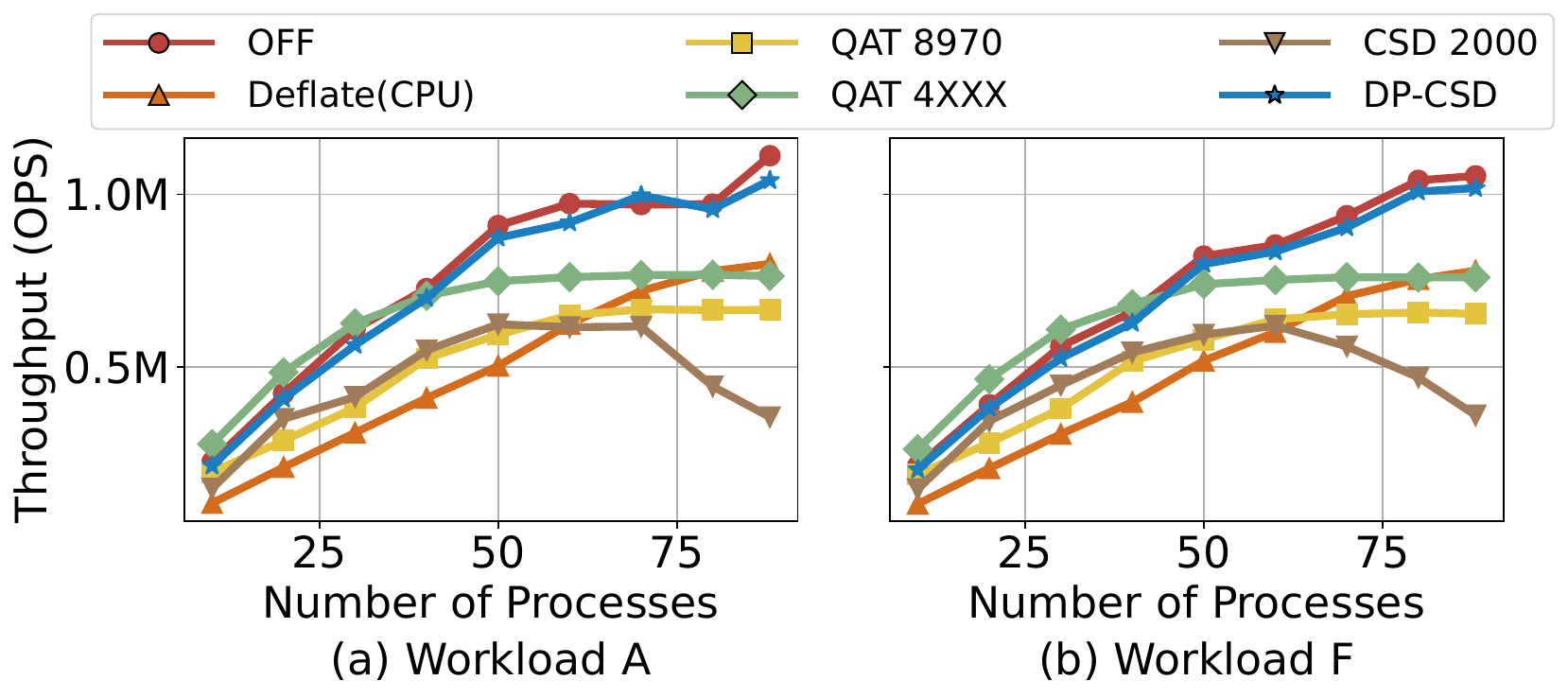} 
    \vspace{-2em}
    \caption{YCSB throughput.}
    \label{fig:ycsb-throughput}
\end{figure}

\subsubsection{RocksDB Database} 
\begin{figure}[t]
    \centering
    \includegraphics[width=0.67\linewidth]{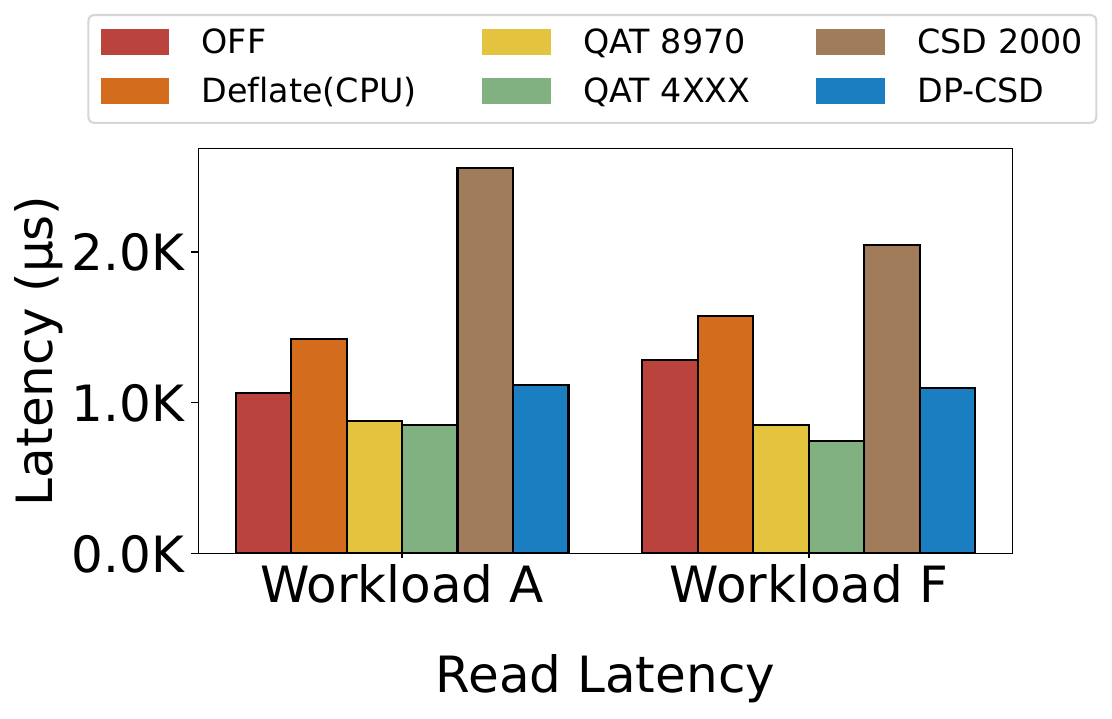} 
    \vspace{-1em}
    \caption{YCSB read latency.}
    \label{fig:ycsb-lat}
\end{figure}

We use YCSB~\cite{ycsb} to evaluate RocksDB, reporting end-to-end throughput (OPS) and latency as concurrency scales (Figures ~\ref{fig:ycsb-throughput} and~\ref{fig:ycsb-lat}). Tested configurations include No Compression (OFF), CPU compression (Deflate), QAT 8970, QAT 4xxx, and DP-CSD with integrated DPZip.
Since we are evaluating data compression, we choose \textit{Workload-A} and \textit{Workload-F}, which target write-intensive and read-modify-write patterns, respectively.

\ul{\textbf{Throughput.}} Baseline RocksDB (\textit{OFF}) delivers the highest throughput in the absence of compression overhead. Even lightweight CPU compression (Deflate level 1) significantly degrades performance: with 10 threads, Workload-A throughput drops 26\% (362$\rightarrow$268 KOPS) and Workload-F by 23.5\% (499$\rightarrow$382 KOPS). In contrast, hardware acceleration with QAT 8970/4xxx not only eliminates this penalty but can also improve throughput by reducing data volume; e.g., QAT 4xxx elevates Workload-A to 476 KOPS at 10 threads. DP-CSD performs almost equivalently to the \textit{OFF} baseline, as in-storage compression does not reduce PCIe data transfer volume.

\textbf{Finding 6. QAT scalability bottlenecks:} \textit{QAT accelerators are limited by hardware queuing and concurrency ceilings (up to 64 concurrent processes), restricting parallelism and causing throughput to plateau.} In contrast, DP-CSD demonstrates superior scalability, reaching 1 MOPS at 88 threads (Workload-F), which is up to 25\% higher than QAT.

\textbf{Finding 7. Vendor-dependent CSD performance:} \textit{CSD performance varies significantly across vendors. While DP-CSD outpaces QAT, legacy CSD 2000 exhibits severe performance degradation under high concurrency, likely due to constrained processing resources and the low-bandwidth (e.g., 2.5GB/s~\cite{chen2024hacsd}) interconnection of its FPGA CDPU within the SSD.}

\ul{\textbf{Latency.}} Due to RocksDB's architecture, where background compaction decouples compression from writes, the effect of different CDPU decompression methods only becomes visible in read operations. To assess true device impact and minimize cache effects, we measured average read latency within 10 seconds after flushing the page cache. 

\textbf{Finding 8. Application-layer compression shapes data layout and latency:} \textit{In RocksDB, QAT-based compression is visible to the application, enabling each SSTable to store more data, thereby reducing LSM tree depth and lowering read latency.} In contrast, host-transparent in-storage compression (e.g., DP-CSD) only reduces the SSTable's physical size on the SSD, but does not change RocksDB's logical organization or tree depth, thus it does not improve read latency. As shown in Figure \ref{fig:ycsb-lat}, QAT-based compression consistently delivers the lowest read latency due to more densely packed SSTables. Although we can increase the SSTable size in the DP-CSD setup to reduce the read latency gap with QAT, doing so also increases compaction overhead in RocksDB. These results highlight that application-level compression can optimize both data layout and performance, but requires tight integration with applications, increasing software complexity and limiting portability across systems.

\subsubsection{\textbf{Filesystems}}
Integrating compression at the filesystem or logical volume manager enables inline compression, as exemplified by QZFS~\cite{qzfs2019} and VDO~\cite{horn2018vdo}. 
We evaluate the performance and energy efficiency of CPU, QAT, and DP-CSD in Btrfs, the only filesystem that currently supports both QAT 8970 and QAT 4xxx accelerators. For end-to-end latency measurements, we use ZFS because it offers flexible configuration of compression block sizes, a feature not available in Btrfs. QAT 4xxx is excluded from ZFS latency evaluation since ZFS currently does not support it.

\begin{figure}[t]
    \centering
    \includegraphics[width=8.5cm]{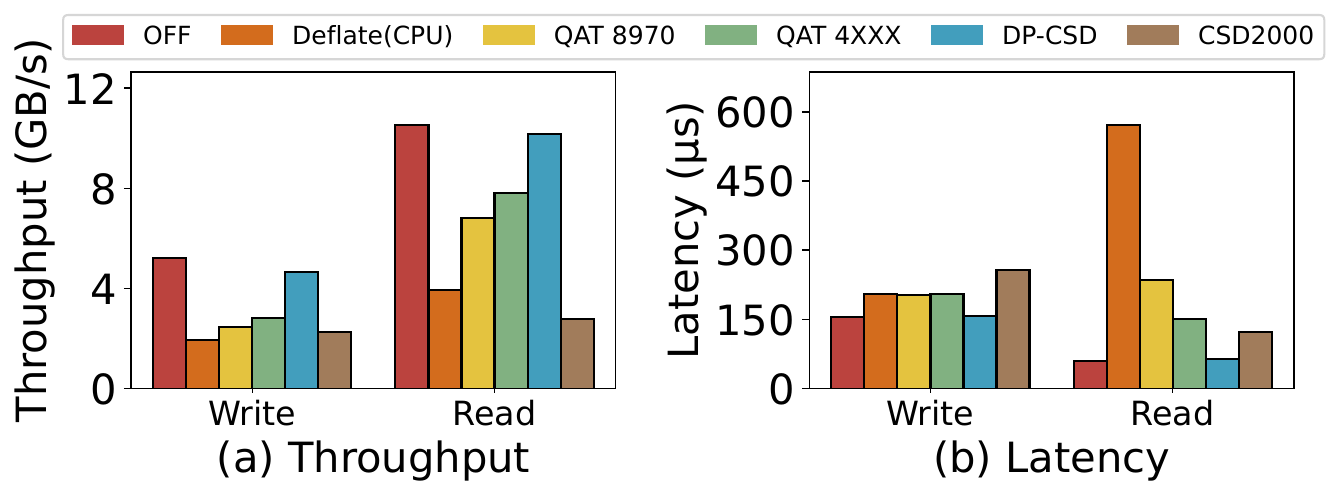}
    \vspace{-2em}
    \caption{Btrfs throughput and latency.}
    \label{fig:btrfs}
\end{figure}

\ul{\textbf{Btrfs Throughput.}}
Btrfs performs asynchronous compression during buffered IO and enforces checksumming when compression is enabled, both of which penalize write bandwidth. As shown in Figure~\ref{fig:btrfs}(a), the highest compression throughput is achieved by DP-CSD. In comparison, QAT 8970 and QAT 4xxx introduce further performance loss with asynchronous compression, while CPU-based Deflate and FPGA-based CSD 2000 compression yield the lowest throughput.
Compression in the QAT-based filesystem layer is based on buffered IO, which introduces additional memory copying overhead and may lead to potential writeback bottlenecks in the background.

\ul{\textbf{Btrfs Latency.}}
Btrfs imposes a maximum compressed block size of 128KB, resulting in significant read amplification: even small reads (e.g., 4KB random accesses) must fetch and decompress the entire 128KB block. As illustrated in Figure~\ref{fig:btrfs}(b), this substantially increases read latency. CPU-based decompression, particularly with Deflate, suffers the most, with latency peaking at 572$\mu$s due to algorithmic inefficiency. QAT devices face similar limitations. Despite the QAT 4xxx's low intrinsic decompression latency, IO stack overheads inflate latency by 90$\mu$s compared to DP-CSD. In contrast, DP-CSD and the \textit{OFF} baseline configuration eliminate read amplification, with DP-CSD introducing only a minimal 5$\mu$s overhead.
Write latency under Btrfs is largely determined by metadata flushing, as buffered IO masks compression costs, resulting in comparable write latencies across all evaluated schemes.

\textbf{Finding 9. Compression block size trade-offs:} \textit{While larger filesystem compression blocks enhance compression ratios, they greatly increase read amplification and random IO latency, especially for CPU-based decompression.}

\begin{figure}[t]
    \centering    
    \includegraphics[height=3.5cm]{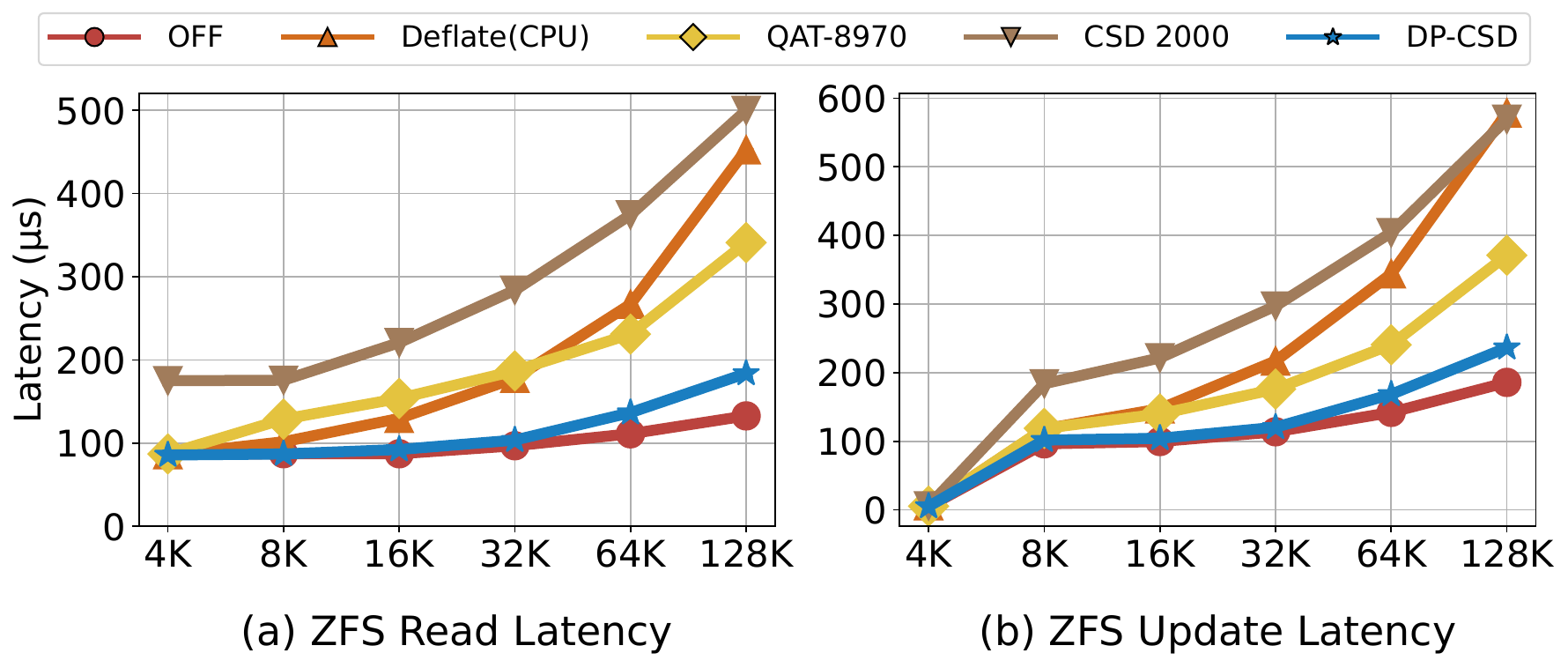}
    \vspace{-1em}
    \caption{ZFS latency.}
    \label{fig:zfs-lat}
\end{figure}

\ul{\textbf{ZFS Filesystem Latency.}}
Figure~\ref{fig:zfs-lat} presents the latency of ZFS across different record sizes (4K to 128K) under various compression methods. CPU-based Deflate compression exhibits the highest latency across all record sizes, with overhead increasing significantly as block size grows, highlighting its inefficiency. QAT 8970 offers only slight improvements over CPU compression and even shows slightly higher latency at smaller block sizes from 8KB to 32KB, likely due to the overhead introduced by the QAT driver's software stack. While its performance is slightly better than CPU compression at larger block sizes, it remains far from optimal.

\textbf{Finding 10. CSD latency and efficiency:} \textit{DP-CSD compression demonstrates consistently lower latency than both CPU and QAT solutions across varying compression block sizes.}
DP-CSD performs only slightly worse than the \textit{OFF} baseline. This indicates that it introduces minimal overhead while maintaining compression benefits. As block size increases, the latency gap between CPU/QAT 8970 and DP-CSD becomes more pronounced, further reinforcing CSD's advantage in delivering high performance with minimal overhead.

\textbf{Finding 11. Filesystem compression overheads:} \textit{Compression in the filesystem (e.g., Btrfs, ZFS, QZFS) layer involves asynchronous handling and additional memory copying, creating bottlenecks such as increased writeback pressure and metadata flushing overheads, which obscure true compression latency and can degrade end-to-end throughput.}

\begin{figure}[t]
    \centering
    \includegraphics[width=8.5cm]{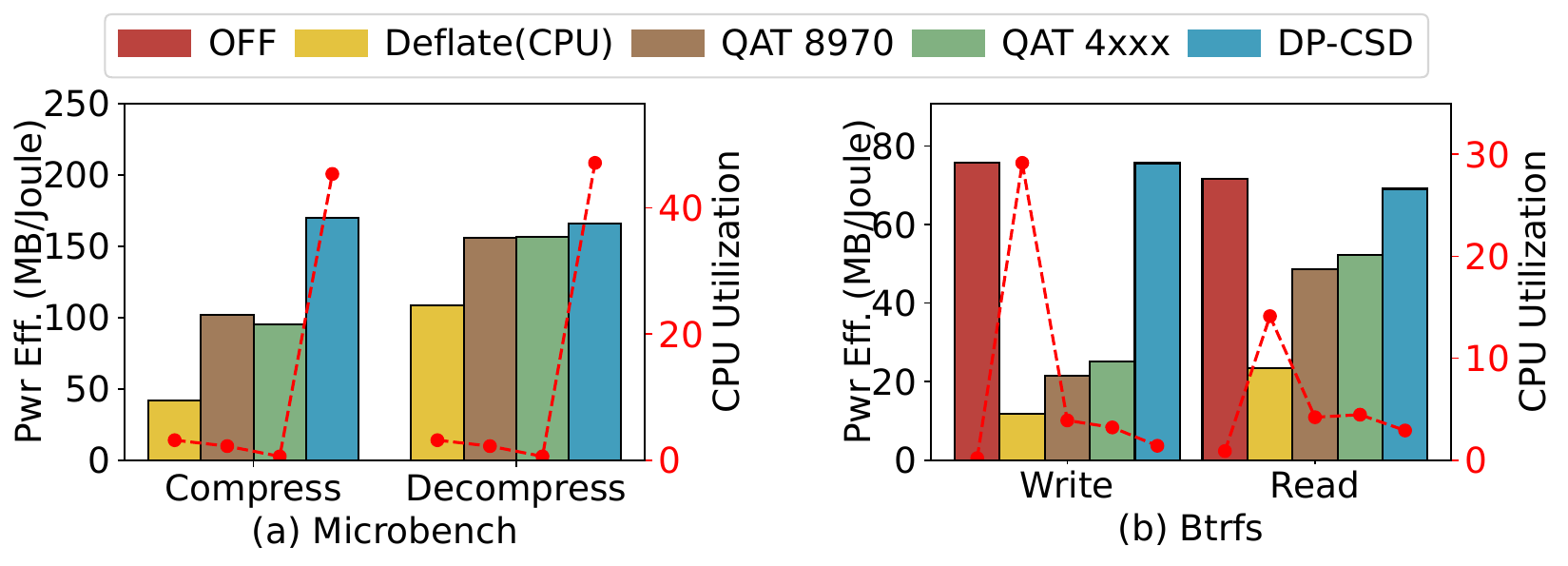}
    \vspace{-2em}
    \caption{Microbench and Btrfs power efficiency.}
    \label{fig:power-eff}
\end{figure}

\subsection{Power Efficiency and TCO}
\subsubsection{Power Efficiency} 
While hardware accelerators dramatically reduce CPU cycles, cutting active energy per data unit by up to two orders of magnitude~\cite{b1_ibmz15_2020}, total system-level energy efficiency is still influenced by data management overheads. Thus, comprehensive evaluation of diverse CDPUs and systems they powered is critical. We collect system-level power data using the out-of-band monitoring capability of server baseboard management controller (BMC). 
To quantify the power efficiency of workloads, we first obtain the net power consumption by subtracting the server's idle power ($P_{\text{idle}}$) from the average workload runtime power ($P_{\text{runtime}}$). Power efficiency is subsequently determined as the ratio of throughput to net power consumption.

\textbf{Finding 12. Non-linear correlation between CDPU and full-system power efficiency:} \textit{CDPU power efficiency may not translate to system power efficiency.}
The DPZip CDPU demonstrates up to $50\times$ higher standalone power efficiency compared to software compression, consuming 2.5W versus 132W for a CPU. However, at the system level, the end-to-end power efficiency improvement decreases to $3.5\times$, as demonstrated by Microbench compression results in Figure~\ref{fig:power-eff}(a).
In a simulated hyperscale environment, both DPZip and QAT CDPUs achieve over 50\% reduction in server electricity costs relative to CPU-based Deflate, while maintaining equivalent throughput. This translates to significant operational savings and reduced CO$_2$ emissions.

\textbf{Finding 13. Power efficiency of DP-CSD and alternatives:}
\textit{DP-CSD delivers the highest power efficiency across device, system, and application levels.} At the device-level, DPZip achieves the best compression (169.87MB/J) and decompression (165.65MB/J) efficiency, outperforming CPU- and QAT-based solutions by 40--45\% (Figure~\ref{fig:power-eff}a). Multi-device scaling further improves DPZip's efficiency to 288.72 MB/J (compression) and 395.88MB/J (decompression), whereas CPU-based Deflate lags at 41.81MB/J. QAT's efficiency suffers due to asynchronous overheads and CPU thread contention.

At the application level using Btrfs, DPZip delivers 75.63 MB/J for write and 69.10 MB/J for read, closely approaching the OFF baseline. Simultaneously, DPZip reduces CPU utilization to less than 3\%, significantly lower than the $>14\%$ observed in software- and QAT-based approaches, which also achieve only 11.75 MB/J for write (Figure~\ref{fig:power-eff}b).

\begin{figure}[t]
    \centering
    \includegraphics[width=8cm]{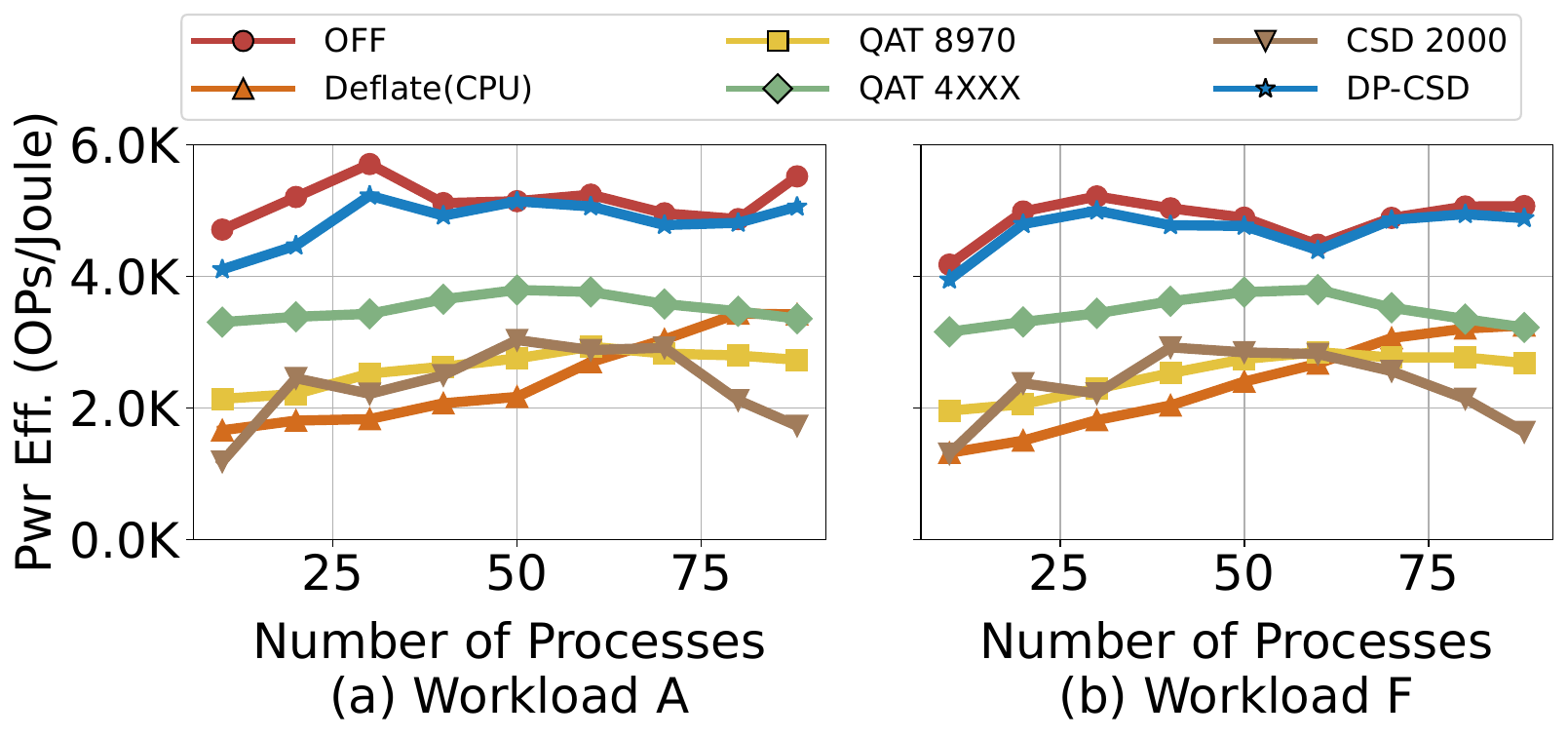}
    \vspace{-1em}
    \caption{YCSB power efficiency.}
    \label{fig:ycsb-poweff}
\end{figure}

For YCSB Workload-A, DPZip achieves superior power efficiency, reaching up to 5224 OPs/J, substantially outperforming both QAT variants, which deliver less than 3800 OPs/J (Figure~\ref{fig:ycsb-poweff}). The limited improvement seen with QAT-based approaches stems primarily from inefficient CPU busy-waiting during hardware polling, resulting in QAT power efficiency comparable to that of software compression.

In summary, DP-CSD demonstrates superior power efficiency compared to CPU and QAT methods, achieving performance metrics that closely match the \textit{OFF} baseline and incurring minimal system overhead.

\subsection{CDPU Scalability and Multi-tenant Sharing}
\subsubsection{Scalability}
Figure~\ref{fig:ycsb-throughput} compares the scalability of various solutions as the thread count increases under a single-accelerator configuration. DP-CSD consistently achieves the highest throughput with increasing threads.
When scaling to multiple accelerators, platform limitations emerge. Both QAT 8970 and DP-CSD scale with the number of available PCIe interfaces, capped at 24 on our server. Unlike DP-CSD, deploying QAT 8970 units induces PCIe port contention and increased hardware costs, further reducing the number of SSDs that can be attached. Similarly, QAT 4xxx scalability is constrained by standard CPU socket counts, typically supporting a maximum of four accelerators per server.

\textbf{Finding 14. Superior scalability of DP-CSD:} \textit{DP-CSD demonstrates linear or near-linear performance scaling with increasing threads or accelerator instances, outperforming QAT, whose scalability is restricted by hardware queue depths (QAT 8970, 4xxx) and CPU integration constraints (QAT 4xxx).} 
On our dual-socket server, QAT 4xxx achieves 4.77GB/s with one accelerator and linearly scales to 9.54GB/s with two. In contrast, a single DP-CSD delivers 12.5GB/s and scales almost linearly to 98.6GB/s with eight devices. Consequently, DP-CSD not only achieves substantially higher throughput than QAT but also enables compression capability to scale proportionally with storage capacity~\cite{wang2025reviving}.

\begin{figure}[t]
    \centering
    \includegraphics[width=8.5cm]{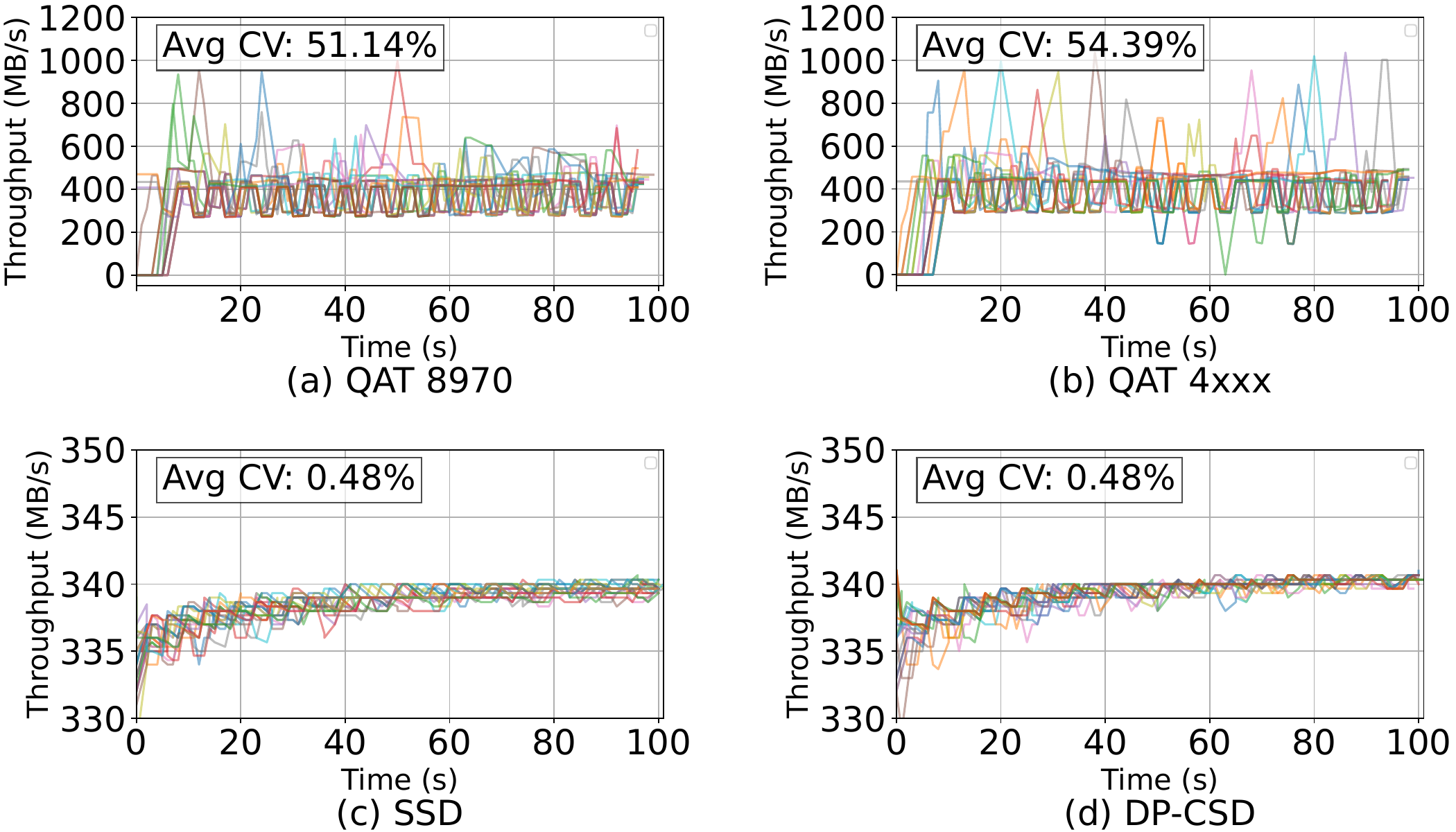}
    \vspace{-2em}
    \caption{Each CDPU is shared by 24 virtual machines via SR-IOV hardware virtualization.}
    \label{fig:vms_throuput_compress}
\end{figure}

\subsubsection{Multi-tenant CDPU Sharing}

We evaluate QoS and multi-tenant isolation using Intel QAT 8970, QAT 4xxx, SSD, and DP-CSD devices fully partitioned into Virtual Functions (VFs). Each VF is exclusively assigned to one of 24 isolated VMs executing independent workloads, enabling precise per-tenant performance measurement under realistic high-concurrency multi-tenant scenarios.

\textbf{Finding 15. Multi-tenant isolation:} \textit{DP-CSD achieves superior multi-VM isolation and throughput stability via SR-IOV, maintaining minimal coefficient of variation ($CV<0.5\%$) compared to QAT-based solutions which exhibit severe performance oscillations ($CV>50\%$) under multi-tenant workloads.}

Multi-VM workload analysis exposes fundamental architectural limitations in QAT-based resource sharing. QAT solutions demonstrate systematic performance degradation with CVs of 54.39\% (QAT 4xxx) and 51.14\% (QAT 8970) for write operations, escalating to 89\% and 80.49\% respectively under read workloads (Figure~\ref{fig:vms_throuput_compress}). This instability stems from the absence of VF isolation mechanisms of QAT devices and severe resource contention among concurrent VM instances. 

In contrast, DP-CSD with DPZip hardware compression achieves stable throughput ($CV=0.48\%$), validating the effectiveness of VF-based resource partitioning for performance isolation among concurrent VM instances. 
Its SR-IOV architecture enforces robust front-end QoS controls, incorporating fair IO scheduling and queue resource management to eliminate resource contention and minimize throughput variability. Consequently, DP-CSD is well-suited for multi-tenant cloud environments that require strong and predictable performance guarantees.

\begin{table}[t]
    \caption{
        CPU-based software compression versus various hardware CDPU solutions. 
    }
    \vspace{-0.5em}
    \centering
    \scriptsize
    \setlength{\tabcolsep}{4pt} 
    \renewcommand{\arraystretch}{1.1} 
    \begin{tabular}{l|c|c|c|c}
        \hline
        & \textbf{CPU} & \textbf{Peripheral}& \textbf{On-chip} & \textbf{In-storage} \\ \hline
        
        \textit{CPU offloading} & \textcolor{red}{\ding{55}} & \textcolor{blue}{\ding{51}} & \textcolor{blue}{\ding{51}} & \textcolor{blue}{\ding{51}}
        \\ \hline

        \textit{Compression acceleration} & \textcolor{red}{\ding{55}} & \textcolor{blue}{\ding{51}}  & \textcolor{blue}{\ding{51}} & \textcolor{blue}{\ding{51}} \\ \hline

        \textit{Cost reduction} & \halfcheck & \halfcheck & \textcolor{blue}{\ding{51}} & \textcolor{blue}{\ding{51}} \\ \hline

        \textit{Power efficiency} & \textcolor{red}{\ding{55}} & \halfcheck & \halfcheck & \textcolor{blue}{\ding{51}} \\ \hline

        \textit{Multi-thread scalability} & \textcolor{blue}{\ding{51}} & \halfcheck  & \halfcheck&\textcolor{blue}{\ding{51}} \\ \hline

        \textit{Multi-device scalability} & \textcolor{red}{\ding{55}} & \textcolor{blue}{\ding{51}}  & \textcolor{red}{\ding{55}}&\textcolor{blue}{\ding{51}} \\ \hline

        \textit{Plug and play} & \textcolor{red}{\ding{55}} & \textcolor{red}{\ding{55}} & \textcolor{red}{\ding{55}} & \textcolor{blue}{\ding{51}} \\ \hline
        \textit{Compression ratio} & \textcolor{blue}{\ding{51}} & \textcolor{blue}{\ding{51}} & \textcolor{blue}{\ding{51}} & \halfcheck  \\ \hline
        \textit{Algorithm configurability} & \textcolor{blue}{\ding{51}} & \halfcheck  & \halfcheck  & \textcolor{red}{\ding{55}} \\ \hline
        
    \end{tabular}
    \label{scheme-summary}
\end{table}

\section{Takeaways and Discussion}\label{discuss}
\textbf{CPU vs. CDPUs.}
Table~\ref{scheme-summary} compares CPU software with three hardware CDPUs. All CDPUs enable CPU offloading and compression acceleration. However, peripheral CDPUs tend to be expensive (e.g., Intel QAT 8970 has an MSRP of \$882~\cite{intel-qat-8970-cdw}), and the high CPU utilization/power overhead of pure software also drives substantial indirect costs. Because peripheral and on-chip CDPUs still require intensive host–device data movement, their energy efficiency lags in-storage CDPUs, which process data directly in IO path. Multi-thread scalability of peripheral and on-chip CDPUs can even trail CPU software due to PCIe request scheduling and limited queue resources. Both software and on-chip approaches scale poorly across multiple devices since capability is bounded by CPU socket counts per server. Moreover, only in-storage CDPUs are genuinely plug-and-play. Other solutions demand heavy application or filesystem adaptation. Indeed, apart from RocksDB and Btrfs, we have not found any other systems that simultaneously support both QAT 8970 and QAT 4xxx series accelerators. Furthermore, even these supported systems accommodate distinct operating systems and software versions individually. In-storage CDPUs face two main limitations: suboptimal compression ratios caused by a fixed 4KB compression granularity and limited algorithm configurability due to tight IO coupling.

\textbf{Architectural diversity and system co-design.}
Our findings indicate that no single CDPU architecture consistently outperforms others across all evaluation criteria. Rather, the three state-of-the-art CDPU architectures exhibit complementary strengths, each excelling in different aspects. Moreover, superior raw CDPU performance does not necessarily yield proportional application performance improvements.

For IO-bound, latency-sensitive workloads, on-chip or in-storage CDPUs provide substantial performance benefits by exploiting memory proximity and minimizing data movement. On-chip CDPUs leverage direct cache-line access, thereby lowering access latency. Conversely, in-storage CDPUs tightly couple to the IO path, effectively eliminating memory copy overhead and maximizing throughput.

\textbf{Workload-aware deployment.}
Architectural choices must be fit for workloads to maximize efficiency. Accurate identification of system-level overheads and bottlenecks requires end-to-end evaluation using application-level benchmarks. When planning for concurrency and scalability, it is critical to account for device-level constraints. CPU-coupled approaches, such as software compression or QAT 4xxx, often encounter resource contention and scaling limitations. Integration layer selection (application, filesystem, or device) should be based on empirical workload characteristics, including concurrency and latency requirements. Compression block sizes must be carefully tuned to minimize performance degradation for small, frequent IOs. Where filesystem-level compression is necessary, design should optimize asynchronous compression execution and efficient metadata management to mitigate performance penalties.

\textbf{CSD deployment recommendations.}
Highly efficient in-storage CDPUs provide optimal power efficiency and throughput. Notably, DP-CSD exhibits linear scalability in compression processing capability with respect to the number of attached CSD devices. This highlights the critical role of computational storage in next-generation architectures, enabling more balanced trade-offs among power, performance, and storage capacity.

\textbf{Remaining R\&D challenges.}
Despite the advances of DPZip and DP‑CSD, several key challenges are to be addressed. First, SSD-native granularities (e.g., 4KB) inherently constrain data redundancy detection, resulting in sub-optimal compression ratios compared to larger block sizes. While preset dictionary compression~\cite{kryczka2021preset} could mitigate this limitation, its design and hardware integration remain open research problems and are earmarked for future work.

Second, DPZip currently supports only a single compression algorithm. Although supporting multiple algorithms could provide flexibility, it would also significantly increase area overhead and design complexity. The current compressor occupies approximately 5\% of controller silicon, and each additional algorithm would further scale this cost. A more scalable and efficient solution is to implement multiple compression levels within a single algorithm, which maintains hardware simplicity while offering tunable performance.

Third, custom hardware compression algorithms face challenges in development costs and ecosystem compatibility. We propose that industry bodies launch standardization efforts akin to the FIPS Cryptographic Module Validation Program~\cite{CMVP}. Establishing unified specifications and a formal certification process will promote cross-vendor interoperability, facilitate coordinated software-hardware design, and streamline product validation. Such standardization not only lowers development barriers and enhances reliability but also accelerates the adoption of CDPU-based technologies, driving sustainable and energy-efficient storage solutions.
Tackling these challenges is crucial for DPZip and future similar CDPU solutions to materialize their full potential.

\section{Conclusion}
CDPUs are advancing rapidly, but their broader system effects have not yet been fully understood. We present the first comprehensive system-level comparative analysis of DPZip and QAT CDPUs, supported by detailed workload profiling, power-performance characterization, and architectural tradeoff evaluation. Our unified analysis of in-storage and on-chip compression accelerators demonstrates that in-storage CDPU architecture offers substantial energy efficiency and scalable throughput for data-centric workloads. However, persistent challenges remain, including sub-optimal compression ratios, limited architectural flexibility, and a lack of ecosystem standardization. By releasing our open methodologies and datasets, we hope to catalyze further research and support the development of standardized, high-performance, and energy-efficient hardware compression solutions.
Beyond CDPU, the distilled design principles and architectural insights may offer generalizable guidelines for optimizing diverse accelerator architectures in other domains.

\begin{acks}
We thank our shepherd, Dr. Javier Gonzalez, the anonymous
reviewers, and Dr. You Zhou for their valuable feedback and suggestions.
This work was supported by the National Key R\&D Program of
China (No. 2023YFB4502901) and
the Shenzhen Key R\&D Program (No. KJZD20240903102459001).
\end{acks}

\bibliographystyle{ACM-Reference-Format}
\bibliography{references}


\begin{thebibliography}{64}


\ifx \showCODEN    \undefined \def \showCODEN     #1{\unskip}     \fi
\ifx \showISBNx    \undefined \def \showISBNx     #1{\unskip}     \fi
\ifx \showISBNxiii \undefined \def \showISBNxiii  #1{\unskip}     \fi
\ifx \showISSN     \undefined \def \showISSN      #1{\unskip}     \fi
\ifx \showLCCN     \undefined \def \showLCCN      #1{\unskip}     \fi
\ifx \shownote     \undefined \def \shownote      #1{#1}          \fi
\ifx \showarticletitle \undefined \def \showarticletitle #1{#1}   \fi
\ifx \showURL      \undefined \def \showURL       {\relax}        \fi
\providecommand\bibfield[2]{#2}
\providecommand\bibinfo[2]{#2}
\providecommand\natexlab[1]{#1}
\providecommand\showeprint[2][]{arXiv:#2}

\bibitem[Abali et~al\mbox{.}(2020)]%
        {b1_ibmz15_2020}
\bibfield{author}{\bibinfo{person}{Bulent Abali}, \bibinfo{person}{Bart
  Blaner}, \bibinfo{person}{John Reilly}, \bibinfo{person}{Matthias Klein},
  \bibinfo{person}{Ashutosh Mishra}, \bibinfo{person}{Craig~B Agricola},
  \bibinfo{person}{Bedri Sendir}, \bibinfo{person}{Alper Buyuktosunoglu},
  \bibinfo{person}{Christian Jacobi}, \bibinfo{person}{William~J Starke},
  {et~al\mbox{.}}} \bibinfo{year}{2020}\natexlab{}.
\newblock \showarticletitle{Data compression accelerator on IBM POWER9 and z15
  processors: Industrial product}. In \bibinfo{booktitle}{\emph{2020 ACM/IEEE
  47th Annual International Symposium on Computer Architecture (ISCA)}}. IEEE,
  \bibinfo{pages}{1--14}.
\newblock


\bibitem[Abdelfattah et~al\mbox{.}(2014)]%
        {abdelfattah2014Gziponachip}
\bibfield{author}{\bibinfo{person}{Mohamed~S. Abdelfattah},
  \bibinfo{person}{Andrei Hagiescu}, {and} \bibinfo{person}{Deshanand Singh}.}
  \bibinfo{year}{2014}\natexlab{}.
\newblock \showarticletitle{Gzip on a Chip: High Performance Lossless Data
  Compression on FPGAs Using OpenCL}. In \bibinfo{booktitle}{\emph{Proceedings
  of the International Workshop on OpenCL (IWOCL)}}. \bibinfo{pages}{4:1--4:9}.
\newblock


\bibitem[{AMD}(2023)]%
        {amd2023maxlinear}
\bibfield{author}{\bibinfo{person}{{AMD}}.} \bibinfo{year}{2023}\natexlab{}.
\newblock \bibinfo{booktitle}{\emph{4th Gen AMD EPYC™ Processors Offer
  Leadership MaxLinear Compression Performance}}.
\newblock \bibinfo{type}{{T}echnical {R}eport}. \bibinfo{institution}{Advanced
  Micro Devices, Inc.}
\newblock
\urldef\tempurl%
\url{https://www.amd.com/content/dam/amd/en/documents/epyc-technical-docs/white-papers/epyc-9004-wp-maxlinear-vs-qat.pdf}
\showURL{%
\tempurl}
\newblock
\shownote{Accessed: 2024-10-25}.


\bibitem[Axboe(2006)]%
        {fio}
\bibfield{author}{\bibinfo{person}{Jens Axboe}.}
  \bibinfo{year}{2006}\natexlab{}.
\newblock \bibinfo{booktitle}{\emph{FIO - Flexible I/O Tester}}.
\newblock
\newblock
\shownote{Available at \url{https://fio.readthedocs.io/}}.


\bibitem[{CDW}(2018)]%
        {intel-qat-8970-cdw}
\bibfield{author}{\bibinfo{person}{{CDW}}.} \bibinfo{year}{2018}\natexlab{}.
\newblock \bibinfo{title}{Intel QuickAssist Adapter 8970 — Cryptographic
  Accelerator — PCIe 3.0 x16}.
\newblock
  \bibinfo{howpublished}{\url{https://www.cdw.com/product/intel-quickassist-adapter-8970-cryptographic-accelerator-pcie-3.0-x16/7572065}}.
\newblock
\newblock
\shownote{Accessed: 2025-05-26}.


\bibitem[Chen et~al\mbox{.}(2021)]%
        {chen2021fpga}
\bibfield{author}{\bibinfo{person}{Jianyu Chen}, \bibinfo{person}{Maurice
  Daverveldt}, {and} \bibinfo{person}{Zaid Al-Ars}.}
  \bibinfo{year}{2021}\natexlab{}.
\newblock \showarticletitle{Fpga acceleration of zstd compression algorithm}.
  In \bibinfo{booktitle}{\emph{2021 IEEE International Parallel and Distributed
  Processing Symposium Workshops (IPDPSW)}}. IEEE, \bibinfo{pages}{188--191}.
\newblock


\bibitem[Chen et~al\mbox{.}(2024)]%
        {chen2024hacsd}
\bibfield{author}{\bibinfo{person}{Xiang Chen}, \bibinfo{person}{Tao Lu},
  \bibinfo{person}{Jiapin Wang}, \bibinfo{person}{Yu Zhong},
  \bibinfo{person}{Guangchun Xie}, \bibinfo{person}{Xueming Cao},
  \bibinfo{person}{Yuanpeng Ma}, {et~al\mbox{.}}}
  \bibinfo{year}{2024}\natexlab{}.
\newblock \showarticletitle{HA-CSD: Host and SSD Coordinated Compression for
  Capacity and Performance}. In \bibinfo{booktitle}{\emph{2024 IEEE
  International Parallel and Distributed Processing Symposium (IPDPS)}}. IEEE,
  \bibinfo{pages}{825--838}.
\newblock


\bibitem[Collet(2011)]%
        {lz4}
\bibfield{author}{\bibinfo{person}{Yann Collet}.}
  \bibinfo{year}{2011}\natexlab{}.
\newblock \bibinfo{title}{LZ4 - Extremely Fast Compression}.
\newblock \bibinfo{howpublished}{\url{https://lz4.github.io/lz4/}}.
\newblock
\newblock
\shownote{Accessed: 2024-11-05}.


\bibitem[Collet(2013)]%
        {collet2013fse}
\bibfield{author}{\bibinfo{person}{Yann Collet}.}
  \bibinfo{year}{2013}\natexlab{}.
\newblock \bibinfo{title}{Finite State Entropy}.
\newblock
  \bibinfo{howpublished}{\url{https://github.com/Cyan4973/FiniteStateEntropy}}.
\newblock
\newblock
\shownote{Accessed: 2023-09-17}.


\bibitem[Collet(2016)]%
        {zstd}
\bibfield{author}{\bibinfo{person}{Yann Collet}.}
  \bibinfo{year}{2016}\natexlab{}.
\newblock \bibinfo{title}{Zstandard - Fast Real-time Compression Algorithm}.
\newblock \bibinfo{howpublished}{\url{https://facebook.github.io/zstd/}}.
\newblock
\newblock
\shownote{Accessed: 2024-11-05}.


\bibitem[Cooper et~al\mbox{.}(2010)]%
        {ycsb}
\bibfield{author}{\bibinfo{person}{Brian~F Cooper}, \bibinfo{person}{Adam
  Silberstein}, \bibinfo{person}{Erwin Tam}, \bibinfo{person}{Raghu
  Ramakrishnan}, {and} \bibinfo{person}{Russell Sears}.}
  \bibinfo{year}{2010}\natexlab{}.
\newblock \showarticletitle{Benchmarking cloud serving systems with YCSB}. In
  \bibinfo{booktitle}{\emph{Proceedings of the 1st ACM symposium on Cloud
  computing}}. \bibinfo{pages}{143--154}.
\newblock


\bibitem[Corporation(2025)]%
        {intel_qatzip}
\bibfield{author}{\bibinfo{person}{Intel Corporation}.}
  \bibinfo{year}{2025}\natexlab{}.
\newblock \bibinfo{title}{QATzip: Compression Library accelerated by
  Intel{\textregistered} QuickAssist Technology}.
\newblock \bibinfo{howpublished}{\url{https://github.com/intel/QATzip}}.
\newblock


\bibitem[Deutsch(1996)]%
        {deutsch1996rfc1951}
\bibfield{author}{\bibinfo{person}{L.~Peter Deutsch}.}
  \bibinfo{year}{1996}\natexlab{}.
\newblock \bibinfo{title}{RFC 1951: DEFLATE Compressed Data Format
  Specification}.
\newblock
  \bibinfo{howpublished}{\url{https://www.rfc-editor.org/info/rfc1951}}.
\newblock
\newblock
\shownote{Accessed: 2024-11-05}.


\bibitem[Deutsch and loup Gailly(1996)]%
        {zlib}
\bibfield{author}{\bibinfo{person}{L.~Peter Deutsch} {and}
  \bibinfo{person}{Jean loup Gailly}.} \bibinfo{year}{1996}\natexlab{}.
\newblock \bibinfo{title}{ZLIB Compressed Data Format Specification version
  3.3}.
\newblock
  \bibinfo{howpublished}{\url{https://www.rfc-editor.org/info/rfc1950}}.
\newblock
\newblock
\shownote{Accessed: 2024-11-05}.


\bibitem[Do et~al\mbox{.}(2020)]%
        {Do2020TOS-Newport}
\bibfield{author}{\bibinfo{person}{Jae~Young Do}, \bibinfo{person}{Victor da
  Cruz~Ferreira}, \bibinfo{person}{Hossein Bobarshad}, \bibinfo{person}{Mahdi
  Torabzadehkashi}, \bibinfo{person}{Siavash Rezaei}, \bibinfo{person}{Ali
  Heydarigorji}, \bibinfo{person}{Diego Fonseca~Pereira de Souza},
  \bibinfo{person}{Brunno~F. Goldstein}, \bibinfo{person}{Leandro Santiago},
  \bibinfo{person}{Min~Soo Kim}, \bibinfo{person}{Priscila M.~V. Lima},
  \bibinfo{person}{Felipe M.~G. Fran{\c{c}}a}, {and}
  \bibinfo{person}{Vladimir~Castro Alves}.} \bibinfo{year}{2020}\natexlab{}.
\newblock \showarticletitle{Cost-Effective, Energy-Efficient, and Scalable
  Storage Computing for Large-Scale {AI} Applications}.
\newblock \bibinfo{journal}{\emph{ACM Transactions on Storage}}
  \bibinfo{volume}{16}, \bibinfo{number}{4} (\bibinfo{year}{2020}),
  \bibinfo{pages}{21:1--21:37}.
\newblock


\bibitem[Dong et~al\mbox{.}(2012)]%
        {dong2012high}
\bibfield{author}{\bibinfo{person}{Yaozu Dong}, \bibinfo{person}{Xiaowei Yang},
  \bibinfo{person}{Jianhui Li}, \bibinfo{person}{Guangdeng Liao},
  \bibinfo{person}{Kun Tian}, {and} \bibinfo{person}{Haibing Guan}.}
  \bibinfo{year}{2012}\natexlab{}.
\newblock \showarticletitle{High performance network virtualization with
  SR-IOV}.
\newblock \bibinfo{journal}{\emph{J. Parallel and Distrib. Comput.}}
  \bibinfo{volume}{72}, \bibinfo{number}{11} (\bibinfo{year}{2012}),
  \bibinfo{pages}{1471--1480}.
\newblock


\bibitem[Duda(2013)]%
        {Duda2013}
\bibfield{author}{\bibinfo{person}{Jarek Duda}.}
  \bibinfo{year}{2013}\natexlab{}.
\newblock \showarticletitle{Asymmetric numeral systems: Entropy coding
  combining speed of Huffman coding with compression rate of arithmetic
  coding}.
\newblock \bibinfo{journal}{\emph{arXiv preprint arXiv:1311.2540}}
  (\bibinfo{year}{2013}).
\newblock
\showeprint[arxiv]{1311.2540}
\urldef\tempurl%
\url{https://arxiv.org/abs/1311.2540}
\showURL{%
\tempurl}


\bibitem[Electronics(2022)]%
        {samsung_smartssd}
\bibfield{author}{\bibinfo{person}{Samsung Electronics}.}
  \bibinfo{year}{2022}\natexlab{}.
\newblock \bibinfo{booktitle}{\emph{Samsung SmartSSD Computational Storage
  Drive}}.
\newblock
\urldef\tempurl%
\url{https://semiconductor.samsung.com/us/ssd/smart-ssd/}
\showURL{%
\tempurl}
\newblock
\shownote{Accessed: 2025-02-11}.


\bibitem[Engineering(2018)]%
        {fb_zstandard_2018}
\bibfield{author}{\bibinfo{person}{Facebook Engineering}.}
  \bibinfo{year}{2018}\natexlab{}.
\newblock \bibinfo{booktitle}{\emph{Zstandard}}.
\newblock
\urldef\tempurl%
\url{https://engineering.fb.com/2018/12/19/core-infra/zstandard/}
\showURL{%
\tempurl}
\newblock
\shownote{Accessed: 2024-02-11}.


\bibitem[{Facebook}(2015)]%
        {facebook_zstd}
\bibfield{author}{\bibinfo{person}{{Facebook}}.}
  \bibinfo{year}{2015}\natexlab{}.
\newblock \bibinfo{title}{{Zstandard: Fast real‐time compression algorithm}}.
\newblock \bibinfo{howpublished}{\url{https://github.com/facebook/zstd}}.
\newblock
\newblock
\shownote{Accessed: 2025-05-11}.


\bibitem[{Facebook Inc.}(2025)]%
        {rocksdb_github}
\bibfield{author}{\bibinfo{person}{{Facebook Inc.}}}
  \bibinfo{year}{2025}\natexlab{}.
\newblock \bibinfo{title}{RocksDB}.
\newblock \bibinfo{howpublished}{\url{https://github.com/facebook/rocksdb}}.
\newblock
\newblock
\shownote{Accessed: 2025-02-11}.


\bibitem[Fakhry et~al\mbox{.}(2023)]%
        {Fakhry2023Array-Review}
\bibfield{author}{\bibinfo{person}{Dina Fakhry}, \bibinfo{person}{Mohamed
  Abdelsalam}, \bibinfo{person}{M~Watheq El-Kharashi}, {and}
  \bibinfo{person}{Mona Safar}.} \bibinfo{year}{2023}\natexlab{}.
\newblock \showarticletitle{A review on computational storage devices and near
  memory computing for high performance applications}.
\newblock \bibinfo{journal}{\emph{Memories-Materials, Devices, Circuits and
  Systems}}  \bibinfo{volume}{4} (\bibinfo{year}{2023}),
  \bibinfo{pages}{100051}.
\newblock


\bibitem[Fowers et~al\mbox{.}(2015)]%
        {fowers2015scalable}
\bibfield{author}{\bibinfo{person}{Jeremy Fowers}, \bibinfo{person}{Joo-Young
  Kim}, \bibinfo{person}{Doug Burger}, {and} \bibinfo{person}{Scott Hauck}.}
  \bibinfo{year}{2015}\natexlab{}.
\newblock \showarticletitle{A scalable high-bandwidth architecture for lossless
  compression on fpgas}. In \bibinfo{booktitle}{\emph{2015 IEEE 23rd Annual
  International Symposium on Field-Programmable Custom Computing Machines}}.
  IEEE, \bibinfo{pages}{52--59}.
\newblock


\bibitem[Gao et~al\mbox{.}(2022)]%
        {gao2022metazip}
\bibfield{author}{\bibinfo{person}{Ruihao Gao}, \bibinfo{person}{Xueqi Li},
  \bibinfo{person}{Yewen Li}, \bibinfo{person}{Xun Wang}, {and}
  \bibinfo{person}{Guangming Tan}.} \bibinfo{year}{2022}\natexlab{}.
\newblock \showarticletitle{MetaZip: A High-throughput and Efficient
  Accelerator for DEFLATE}. In \bibinfo{booktitle}{\emph{Proceedings of the
  59th ACM/IEEE Design Automation Conference (DAC)}}.
  \bibinfo{pages}{319--324}.
\newblock


\bibitem[Gao et~al\mbox{.}(2024)]%
        {beezip2024asplos}
\bibfield{author}{\bibinfo{person}{Ruihao Gao}, \bibinfo{person}{Zhichun Li},
  \bibinfo{person}{Guangming Tan}, {and} \bibinfo{person}{Xueqi Li}.}
  \bibinfo{year}{2024}\natexlab{}.
\newblock \showarticletitle{Beezip: Towards an organized and scalable
  architecture for data compression}. In \bibinfo{booktitle}{\emph{Proceedings
  of the 29th ACM International Conference on Architectural Support for
  Programming Languages and Operating Systems, Volume 3}}.
  \bibinfo{pages}{133--148}.
\newblock


\bibitem[Gonzalez et~al\mbox{.}(2023)]%
        {gonzalez2023profiling}
\bibfield{author}{\bibinfo{person}{Abraham Gonzalez}, \bibinfo{person}{Aasheesh
  Kolli}, \bibinfo{person}{Samira Khan}, \bibinfo{person}{Sihang Liu},
  \bibinfo{person}{Vidushi Dadu}, \bibinfo{person}{Sagar Karandikar},
  \bibinfo{person}{Jichuan Chang}, \bibinfo{person}{Krste Asanovic}, {and}
  \bibinfo{person}{Parthasarathy Ranganathan}.}
  \bibinfo{year}{2023}\natexlab{}.
\newblock \showarticletitle{Profiling hyperscale big data processing}. In
  \bibinfo{booktitle}{\emph{Proceedings of the 50th Annual International
  Symposium on Computer Architecture}}. \bibinfo{pages}{1--16}.
\newblock


\bibitem[Google(2011)]%
        {snappy}
\bibfield{author}{\bibinfo{person}{Google}.} \bibinfo{year}{2011}\natexlab{}.
\newblock \bibinfo{title}{Snappy: A Fast Compressor/Decompressor}.
\newblock \bibinfo{howpublished}{\url{https://github.com/google/snappy}}.
\newblock
\newblock
\shownote{Accessed: 2024-11-05}.


\bibitem[Google(2023)]%
        {HyperCompressBench}
\bibfield{author}{\bibinfo{person}{Google}.} \bibinfo{year}{2023}\natexlab{}.
\newblock \bibinfo{title}{HyperCompressBench}.
\newblock
  \bibinfo{howpublished}{\url{https://github.com/google/HyperCompressBench}}.
\newblock


\bibitem[Heydarigorji et~al\mbox{.}(2022)]%
        {Heydarigorji2022TECS-CSD}
\bibfield{author}{\bibinfo{person}{Ali Heydarigorji}, \bibinfo{person}{Siavash
  Rezaei}, \bibinfo{person}{Mahdi Torabzadehkashi}, \bibinfo{person}{Hossein
  Bobarshad}, \bibinfo{person}{Vladimir~Castro Alves}, {and}
  \bibinfo{person}{Pai~H. Chou}.} \bibinfo{year}{2022}\natexlab{}.
\newblock \showarticletitle{Leveraging Computational Storage for
  Power-Efficient Distributed Data Analytics}.
\newblock \bibinfo{journal}{\emph{ACM Transactions on Embedded Computing
  Systems}} \bibinfo{volume}{21}, \bibinfo{number}{6} (\bibinfo{year}{2022}),
  \bibinfo{pages}{82:1--82:36}.
\newblock


\bibitem[Horn(2018)]%
        {horn2018vdo}
\bibfield{author}{\bibinfo{person}{Christian Horn}.}
  \bibinfo{year}{2018}\natexlab{}.
\newblock \bibinfo{title}{A Look at VDO, the New Linux Compression Layer}.
\newblock
  \bibinfo{howpublished}{\url{https://www.redhat.com/en/blog/look-vdo-new-linux-compression-layer}}.
\newblock
\newblock
\shownote{Accessed: 2024-11-05}.


\bibitem[Hu et~al\mbox{.}(2019)]%
        {qzfs2019}
\bibfield{author}{\bibinfo{person}{Xiaokang Hu}, \bibinfo{person}{Fuzong Wang},
  \bibinfo{person}{Weigang Li}, \bibinfo{person}{Jian Li}, {and}
  \bibinfo{person}{Haibing Guan}.} \bibinfo{year}{2019}\natexlab{}.
\newblock \showarticletitle{QZFS: QAT Accelerated Compression in File System
  for Application Agnostic and Cost Efficient Data Storage}. In
  \bibinfo{booktitle}{\emph{2019 USENIX Annual Technical Conference (USENIX ATC
  19)}}. \bibinfo{pages}{163--176}.
\newblock


\bibitem[Huffman(1952)]%
        {huffman1952method}
\bibfield{author}{\bibinfo{person}{David~A. Huffman}.}
  \bibinfo{year}{1952}\natexlab{}.
\newblock \showarticletitle{A method for the construction of minimum-redundancy
  codes}.
\newblock \bibinfo{journal}{\emph{Proceedings of the IRE}}
  \bibinfo{volume}{40}, \bibinfo{number}{9} (\bibinfo{year}{1952}),
  \bibinfo{pages}{1098--1101}.
\newblock


\bibitem[{IBM Corporation}(2015)]%
        {ibm_zedc}
\bibfield{author}{\bibinfo{person}{{IBM Corporation}}.}
  \bibinfo{year}{2015}\natexlab{}.
\newblock \bibinfo{booktitle}{\emph{IBM zEDC}}.
\newblock IBM Corporation.
\newblock
\urldef\tempurl%
\url{https://www.ibm.com/docs/en/zos/3.1.0?topic=zdcz-overview-planning-zenterprise-data-compression-zedc}
\showURL{%
\tempurl}
\newblock
\shownote{Product documentation. Accessed: 2024-02-11}.


\bibitem[{Intel Corporation}(2015)]%
        {Intel-DDIO}
\bibfield{author}{\bibinfo{person}{{Intel Corporation}}.}
  \bibinfo{year}{2015}\natexlab{}.
\newblock \bibinfo{title}{Intel\textregistered{} Data Direct I/O Technology
  (Intel\textregistered{} DDIO)}.
\newblock
\urldef\tempurl%
\url{https://www.intel.com/content/www/us/en/io/data-direct-i-o-technology.html}
\showURL{%
\tempurl}
\newblock
\shownote{Accessed: 2025-08-21}.


\bibitem[{Intel Corporation.}(2024)]%
        {website:qat}
\bibfield{author}{\bibinfo{person}{{Intel Corporation.}}}
  \bibinfo{year}{2024}\natexlab{}.
\newblock \bibinfo{title}{{Specification of Intel® QuickAssist Adapter 8970}}.
\newblock
  \bibinfo{howpublished}{\url{https://www.intel.com/content/www/us/en/products/sku/125200/intel-quickassist-adapter-8970/specifications.html}}.
\newblock
\newblock
\shownote{Accessed: 2025-05-26}.


\bibitem[Jun et~al\mbox{.}(2015)]%
        {Jun2015ISCA-BlueDBM}
\bibfield{author}{\bibinfo{person}{Sang{-}Woo Jun}, \bibinfo{person}{Ming Liu},
  \bibinfo{person}{Sungjin Lee}, \bibinfo{person}{Jamey Hicks},
  \bibinfo{person}{John Ankcorn}, \bibinfo{person}{Myron King},
  \bibinfo{person}{Shuotao Xu}, \bibinfo{person}{Arvind}, {and}
  \bibinfo{person}{Srinivas Devadas}.} \bibinfo{year}{2015}\natexlab{}.
\newblock \showarticletitle{BlueDBM: An Appliance for Big Data Analytics}. In
  \bibinfo{booktitle}{\emph{Proceedings of the 42nd Annual International
  Symposium on Computer Architecture (ISCA)}}.
\newblock


\bibitem[Kanev et~al\mbox{.}(2015)]%
        {kanev2015profiling}
\bibfield{author}{\bibinfo{person}{Svilen Kanev}, \bibinfo{person}{Juan~Pablo
  Darago}, \bibinfo{person}{Kim Hazelwood}, \bibinfo{person}{Parthasarathy
  Ranganathan}, \bibinfo{person}{Tipp Moseley}, \bibinfo{person}{Gu-Yeon Wei},
  {and} \bibinfo{person}{David Brooks}.} \bibinfo{year}{2015}\natexlab{}.
\newblock \showarticletitle{Profiling a warehouse-scale computer}. In
  \bibinfo{booktitle}{\emph{Proceedings of the 42nd annual international
  symposium on computer architecture}}. \bibinfo{pages}{158--169}.
\newblock


\bibitem[Karandikar et~al\mbox{.}(2023)]%
        {karandikar2023cdpu}
\bibfield{author}{\bibinfo{person}{Sagar Karandikar},
  \bibinfo{person}{Aniruddha~N. Udipi}, \bibinfo{person}{Junsun Choi},
  \bibinfo{person}{Joonho Whangbo}, \bibinfo{person}{Jerry Zhao},
  \bibinfo{person}{Svilen Kanev}, \bibinfo{person}{Edwin Lim}, {et~al\mbox{.}}}
  \bibinfo{year}{2023}\natexlab{}.
\newblock \showarticletitle{CDPU: Co-designing Compression and Decompression
  Processing Units for Hyperscale Systems}. In
  \bibinfo{booktitle}{\emph{Proceedings of the 50th Annual International
  Symposium on Computer Architecture}}. \bibinfo{pages}{1--17}.
\newblock


\bibitem[Kryczka(2021)]%
        {kryczka2021preset}
\bibfield{author}{\bibinfo{person}{Andrew Kryczka}.}
  \bibinfo{year}{2021}\natexlab{}.
\newblock \bibinfo{title}{Preset Dictionary Compression}.
\newblock
\urldef\tempurl%
\url{https://rocksdb.org/blog/2021/05/31/dictionary-compression.html}
\showURL{%
\tempurl}
\newblock
\shownote{Accessed: 2025-08-29}.


\bibitem[Lazarev et~al\mbox{.}(2024)]%
        {lazarev2024sabre}
\bibfield{author}{\bibinfo{person}{Nikita Lazarev}, \bibinfo{person}{Varun
  Gohil}, \bibinfo{person}{James Tsai}, \bibinfo{person}{Andy Anderson},
  \bibinfo{person}{Bhushan Chitlur}, \bibinfo{person}{Zhiru Zhang}, {and}
  \bibinfo{person}{Christina Delimitrou}.} \bibinfo{year}{2024}\natexlab{}.
\newblock \showarticletitle{Sabre: {Hardware-Accelerated} Snapshot Compression
  for Serverless {MicroVMs}}. In \bibinfo{booktitle}{\emph{18th USENIX
  Symposium on Operating Systems Design and Implementation (OSDI 24)}}. USENIX,
  \bibinfo{pages}{1--18}.
\newblock


\bibitem[Li et~al\mbox{.}(2023)]%
        {li2023more}
\bibfield{author}{\bibinfo{person}{Qiang Li}, \bibinfo{person}{Qiao Xiang},
  \bibinfo{person}{Yuxin Wang}, \bibinfo{person}{Haohao Song},
  \bibinfo{person}{Ridi Wen}, \bibinfo{person}{Wenhui Yao},
  \bibinfo{person}{Yuanyuan Dong}, \bibinfo{person}{Shuqi Zhao},
  \bibinfo{person}{Shuo Huang}, \bibinfo{person}{Zhaosheng Zhu},
  {et~al\mbox{.}}} \bibinfo{year}{2023}\natexlab{}.
\newblock \showarticletitle{More than capacity: Performance-oriented evolution
  of pangu in alibaba}. In \bibinfo{booktitle}{\emph{21st USENIX Conference on
  File and Storage Technologies (FAST 23)}}. \bibinfo{pages}{331--346}.
\newblock


\bibitem[Li et~al\mbox{.}(2025)]%
        {StreamCSD}
\bibfield{author}{\bibinfo{person}{Wenjie Li}, \bibinfo{person}{Xiang Chen},
  \bibinfo{person}{Yelin Shan}, \bibinfo{person}{Jiapin Wang},
  \bibinfo{person}{Yunxin Huang}, \bibinfo{person}{Yafei Yang},
  \bibinfo{person}{Tao Lu}, \bibinfo{person}{You Zhou}, {and}
  \bibinfo{person}{Fei Wu}.} \bibinfo{year}{2025}\natexlab{}.
\newblock \showarticletitle{StreamCSD: SSD-Autonomous Stream Management via
  In-Storage Content Learning}. In \bibinfo{booktitle}{\emph{2025 62nd ACM/IEEE
  Design Automation Conference (DAC)}}. \bibinfo{pages}{1--7}.
\newblock


\bibitem[Lou et~al\mbox{.}(2025)]%
        {lou2025dynamic}
\bibfield{author}{\bibinfo{person}{Jiaqi Lou}, \bibinfo{person}{Srikar
  Vanavasam}, \bibinfo{person}{Yifan Yuan}, \bibinfo{person}{Ren Wang}, {and}
  \bibinfo{person}{Nam~Sung Kim}.} \bibinfo{year}{2025}\natexlab{}.
\newblock \showarticletitle{Dynamic Load Balancer in Intel Xeon Scalable
  Processor: Performance Analyses, Enhancements, and Guidelines}. In
  \bibinfo{booktitle}{\emph{Proceedings of the 52nd Annual International
  Symposium on Computer Architecture}}. \bibinfo{pages}{664--678}.
\newblock


\bibitem[loup Gailly and Adler(1992)]%
        {gzip}
\bibfield{author}{\bibinfo{person}{Jean loup Gailly} {and}
  \bibinfo{person}{Mark Adler}.} \bibinfo{year}{1992}\natexlab{}.
\newblock \bibinfo{title}{Gzip}.
\newblock \bibinfo{howpublished}{\url{https://www.gnu.org/software/gzip/}}.
\newblock
\newblock
\shownote{Accessed: 2024-11-05}.


\bibitem[Mahapatra et~al\mbox{.}(2025)]%
        {mahapatra2025storage}
\bibfield{author}{\bibinfo{person}{Rohan Mahapatra}, \bibinfo{person}{Harsha
  Santhanam}, \bibinfo{person}{Christopher Priebe}, \bibinfo{person}{Hanyang
  Xu}, {and} \bibinfo{person}{Hadi Esmaeilzadeh}.}
  \bibinfo{year}{2025}\natexlab{}.
\newblock \showarticletitle{In-Storage Acceleration of Retrieval Augmented
  Generation as a Service}. In \bibinfo{booktitle}{\emph{Proceedings of the
  52nd Annual International Symposium on Computer Architecture}}.
  \bibinfo{pages}{450--466}.
\newblock


\bibitem[{NVM Express Workgroup}(2025)]%
        {nvme2025}
\bibfield{author}{\bibinfo{person}{{NVM Express Workgroup}}.}
  \bibinfo{year}{2025}\natexlab{}.
\newblock \bibinfo{title}{NVM Express Base Specification, Revision 2.3}.
\newblock
  \bibinfo{howpublished}{\url{https://nvmexpress.org/wp-content/uploads/NVM-Express-Base-Specification-Revision-2.3-2025.08.01-Ratified.pdf}}.
\newblock
\newblock
\shownote{Accessed: 2025-09-26}.


\bibitem[of~Standards and Technology(2025)]%
        {CMVP}
\bibfield{author}{\bibinfo{person}{National~Institute of Standards} {and}
  \bibinfo{person}{Technology}.} \bibinfo{year}{2025}\natexlab{}.
\newblock \bibinfo{title}{Cryptographic Module Validation Program}.
\newblock
\urldef\tempurl%
\url{https://csrc.nist.gov/projects/cryptographic-module-validation-program}
\showURL{%
\tempurl}
\newblock
\shownote{Accessed: 2025-09-20}.


\bibitem[of~Technology(2003)]%
        {silesia_corpus}
\bibfield{author}{\bibinfo{person}{Silesian~University of Technology}.}
  \bibinfo{year}{2003}\natexlab{}.
\newblock \bibinfo{title}{Silesia Compression Corpus}.
\newblock
  \bibinfo{howpublished}{\url{https://sun.aei.polsl.pl//~sdeor/index.php?page=silesia}}.
\newblock
\newblock
\shownote{Accessed: 2024-10-17}.


\bibitem[Okafor et~al\mbox{.}(2024)]%
        {Okafor2024JETC-Fuse}
\bibfield{author}{\bibinfo{person}{Ikenna Okafor},
  \bibinfo{person}{Akshay~Krishna Ramanathan}, \bibinfo{person}{Nagadastagiri
  Challapalle}, \bibinfo{person}{Zheyu Li}, {and}
  \bibinfo{person}{Vijaykrishnan Narayanan}.} \bibinfo{year}{2024}\natexlab{}.
\newblock \showarticletitle{Fusing In-Storage and Near-Storage Acceleration of
  Convolutional Neural Networks}.
\newblock \bibinfo{journal}{\emph{ACM Journal on Emerging Technologies in
  Computing Systems}} \bibinfo{volume}{20}, \bibinfo{number}{1}
  (\bibinfo{year}{2024}), \bibinfo{pages}{1:1--1:22}.
\newblock


\bibitem[{Open Compute Project}(2019)]%
        {microsoft2019zipline}
\bibfield{author}{\bibinfo{person}{{Open Compute Project}}.}
  \bibinfo{year}{2019}\natexlab{}.
\newblock \bibinfo{title}{Project Zipline}.
\newblock
  \bibinfo{howpublished}{\url{https://github.com/opencomputeproject/Project-Zipline}}.
\newblock
\newblock
\shownote{Accessed: 2025-02-09}.


\bibitem[Qiao et~al\mbox{.}(2018)]%
        {qiao2018high}
\bibfield{author}{\bibinfo{person}{Weikang Qiao}, \bibinfo{person}{Jieqiong
  Du}, \bibinfo{person}{Zhenman Fang}, \bibinfo{person}{Michael Lo},
  \bibinfo{person}{Mau-Chung~Frank Chang}, {and} \bibinfo{person}{Jason Cong}.}
  \bibinfo{year}{2018}\natexlab{}.
\newblock \showarticletitle{High-throughput lossless compression on tightly
  coupled CPU-FPGA platforms}. In \bibinfo{booktitle}{\emph{2018 IEEE 26th
  Annual International Symposium on Field-Programmable Custom Computing
  Machines (FCCM)}}. IEEE, \bibinfo{pages}{37--44}.
\newblock


\bibitem[Qiao et~al\mbox{.}(2022)]%
        {b5_csd2000}
\bibfield{author}{\bibinfo{person}{Yifan Qiao}, \bibinfo{person}{Xubin Chen},
  \bibinfo{person}{Ning Zheng}, \bibinfo{person}{Jiangpeng Li},
  \bibinfo{person}{Yang Liu}, {and} \bibinfo{person}{Tong Zhang}.}
  \bibinfo{year}{2022}\natexlab{}.
\newblock \showarticletitle{Closing the B+-tree vs. LSM-tree Write
  Amplification Gap on Modern Storage Hardware with Built-in Transparent
  Compression}. In \bibinfo{booktitle}{\emph{20th USENIX Conference on File and
  Storage Technologies (FAST 22)}}. \bibinfo{pages}{69--82}.
\newblock


\bibitem[Rodeh et~al\mbox{.}(2013)]%
        {rodeh2013btrfs}
\bibfield{author}{\bibinfo{person}{Ohad Rodeh}, \bibinfo{person}{Josef Bacik},
  {and} \bibinfo{person}{Chris Mason}.} \bibinfo{year}{2013}\natexlab{}.
\newblock \showarticletitle{BTRFS: The Linux B-tree filesystem}.
\newblock \bibinfo{journal}{\emph{ACM Transactions on Storage (TOS)}}
  \bibinfo{volume}{9}, \bibinfo{number}{3} (\bibinfo{year}{2013}),
  \bibinfo{pages}{1--32}.
\newblock


\bibitem[Salamat et~al\mbox{.}(2021)]%
        {Salamat2021FPGA-NASCENT}
\bibfield{author}{\bibinfo{person}{Sahand Salamat}, \bibinfo{person}{Armin~Haj
  Aboutalebi}, \bibinfo{person}{Behnam Khaleghi}, \bibinfo{person}{Joo~Hwan
  Lee}, \bibinfo{person}{Yang~Seok Ki}, {and} \bibinfo{person}{Tajana~S.
  Rosing}.} \bibinfo{year}{2021}\natexlab{}.
\newblock \showarticletitle{NASCENT: Near-Storage Acceleration of Database Sort
  on SmartSSD}. In \bibinfo{booktitle}{\emph{FPGA '21: Proceedings of the 2021
  ACM/SIGDA International Symposium on Field-Programmable Gate Arrays}}.
  \bibinfo{pages}{262--272}.
\newblock


\bibitem[Satyanarayanan et~al\mbox{.}(2022)]%
        {Satyanarayanan2022Micro-SilentWitness}
\bibfield{author}{\bibinfo{person}{Mahadev Satyanarayanan},
  \bibinfo{person}{Ziqiang Feng}, \bibinfo{person}{Shilpa George},
  \bibinfo{person}{Jan Harkes}, \bibinfo{person}{Roger Iyengar},
  \bibinfo{person}{Haithem Turki}, {and} \bibinfo{person}{Padmanabhan Pillai}.}
  \bibinfo{year}{2022}\natexlab{}.
\newblock \showarticletitle{Accelerating Silent Witness Storage}.
\newblock \bibinfo{journal}{\emph{IEEE Micro}} \bibinfo{volume}{42},
  \bibinfo{number}{6} (\bibinfo{year}{2022}), \bibinfo{pages}{39--47}.
\newblock


\bibitem[{ScaleFlux, Inc.}(2022)]%
        {scaleflux_csd2000}
\bibfield{author}{\bibinfo{person}{{ScaleFlux, Inc.}}}
  \bibinfo{year}{2022}\natexlab{}.
\newblock \bibinfo{title}{CSD 2000 Series Computational Storage Drive}.
\newblock
  \bibinfo{howpublished}{\url{https://scaleflux.com/products/csd-2000/}}.
\newblock
\newblock
\shownote{Accessed: 2025-04-16}.


\bibitem[SNIA(2022)]%
        {snia_csd_v1}
\bibfield{author}{\bibinfo{person}{SNIA}.} \bibinfo{year}{2022}\natexlab{}.
\newblock \bibinfo{title}{Computational Storage Architecture and Programming
  Model Version 1.0}.
\newblock
  \bibinfo{howpublished}{\url{https://www.snia.org/sites/default/files/technical-work/computational/release/SNIA-Computational-Storage-Architecture-and-Programming-Model-1.0.pdf}}.
\newblock
\newblock
\shownote{Online; accessed on September 3, 2022}.


\bibitem[Sriraman and Dhanotia(2020)]%
        {sriraman2020accelerometer}
\bibfield{author}{\bibinfo{person}{Akshitha Sriraman} {and}
  \bibinfo{person}{Abhishek Dhanotia}.} \bibinfo{year}{2020}\natexlab{}.
\newblock \showarticletitle{Accelerometer: Understanding acceleration
  opportunities for data center overheads at hyperscale}. In
  \bibinfo{booktitle}{\emph{Proceedings of the Twenty-Fifth International
  Conference on Architectural Support for Programming Languages and Operating
  Systems}}. \bibinfo{pages}{733--750}.
\newblock


\bibitem[Tian et~al\mbox{.}(2024)]%
        {tian2024scalable}
\bibfield{author}{\bibinfo{person}{Bing Tian}, \bibinfo{person}{Haikun Liu},
  \bibinfo{person}{Zhuohui Duan}, \bibinfo{person}{Xiaofei Liao},
  \bibinfo{person}{Hai Jin}, {and} \bibinfo{person}{Yu Zhang}.}
  \bibinfo{year}{2024}\natexlab{}.
\newblock \showarticletitle{Scalable Billion-point Approximate Nearest Neighbor
  Search Using SmartSSDs}. In \bibinfo{booktitle}{\emph{2024 USENIX Annual
  Technical Conference (USENIX ATC 24)}}. \bibinfo{pages}{1135--1150}.
\newblock
\urldef\tempurl%
\url{https://www.usenix.org/conference/atc24/presentation/tian}
\showURL{%
\tempurl}


\bibitem[Wang et~al\mbox{.}(2025)]%
        {wang2025reviving}
\bibfield{author}{\bibinfo{person}{Yingjia Wang}, \bibinfo{person}{Tao Lu},
  \bibinfo{person}{Yuhong Liang}, \bibinfo{person}{Xiang Chen}, {and}
  \bibinfo{person}{Ming-Chang Yang}.} \bibinfo{year}{2025}\natexlab{}.
\newblock \showarticletitle{Reviving In-Storage Hardware Compression on ZNS
  SSDs through Host-SSD Collaboration}. In \bibinfo{booktitle}{\emph{2025 IEEE
  International Symposium on High Performance Computer Architecture (HPCA)}}.
  IEEE, \bibinfo{pages}{608--623}.
\newblock


\bibitem[Yuan et~al\mbox{.}(2024)]%
        {yuan2024intel}
\bibfield{author}{\bibinfo{person}{Yifan Yuan}, \bibinfo{person}{Ren Wang},
  \bibinfo{person}{Narayan Ranganathan}, \bibinfo{person}{Nikhil Rao},
  \bibinfo{person}{Sanjay Kumar}, \bibinfo{person}{Philip Lantz},
  \bibinfo{person}{Vivekananthan Sanjeepan}, {et~al\mbox{.}}}
  \bibinfo{year}{2024}\natexlab{}.
\newblock \showarticletitle{Intel Accelerators Ecosystem: An SoC-Oriented
  Perspective: Industry Product}. In \bibinfo{booktitle}{\emph{2024 ACM/IEEE
  51st Annual International Symposium on Computer Architecture (ISCA)}}. IEEE,
  \bibinfo{pages}{848--862}.
\newblock


\bibitem[Zheng et~al\mbox{.}(2022)]%
        {Zheng2022LCTES-ISKEVA}
\bibfield{author}{\bibinfo{person}{Yi Zheng}, \bibinfo{person}{Joshua Fixelle},
  \bibinfo{person}{Nagadastagiri Challapalle}, \bibinfo{person}{Pingyi Huo},
  \bibinfo{person}{Zhaoyan Shen}, \bibinfo{person}{Zili Shao},
  \bibinfo{person}{Mircea~R. Stan}, {and} \bibinfo{person}{Vijaykrishnan
  Narayanan}.} \bibinfo{year}{2022}\natexlab{}.
\newblock \showarticletitle{ISKEVA: In-SSD Key-Value Database Engine for Video
  Analytics Applications}. In \bibinfo{booktitle}{\emph{Proceedings of the 23rd
  ACM SIGPLAN/SIGBED International Conference on Languages, Compilers, and
  Tools for Embedded Systems (LCTES '22)}}.
\newblock


\bibitem[Zhou et~al\mbox{.}(2024)]%
        {zhou2024csal}
\bibfield{author}{\bibinfo{person}{Yanbo Zhou}, \bibinfo{person}{Erci Xu},
  \bibinfo{person}{Li Zhang}, \bibinfo{person}{Kapil Karkra},
  \bibinfo{person}{Mariusz Barczak}, \bibinfo{person}{Wayne Gao},
  \bibinfo{person}{Wojciech Malikowski}, \bibinfo{person}{Mateusz Kozlowski},
  \bibinfo{person}{\L{}ukasz \L{}asek}, \bibinfo{person}{Ruiming Lu},
  \bibinfo{person}{Feng Yang}, \bibinfo{person}{Lilong Huang},
  \bibinfo{person}{Xiaolu Zhang}, \bibinfo{person}{Keqiang Niu},
  \bibinfo{person}{Jiaji Zhu}, {and} \bibinfo{person}{Jiesheng Wu}.}
  \bibinfo{year}{2024}\natexlab{}.
\newblock \showarticletitle{CSAL: the Next-Gen Local Disks for the Cloud}. In
  \bibinfo{booktitle}{\emph{Proceedings of the Nineteenth European Conference
  on Computer Systems}} (Athens, Greece) \emph{(\bibinfo{series}{EuroSys
  '24})}. \bibinfo{pages}{608–623}.
\newblock


\bibitem[Ziv and Lempel(1977)]%
        {ziv1977universal}
\bibfield{author}{\bibinfo{person}{Jacob Ziv} {and} \bibinfo{person}{Abraham
  Lempel}.} \bibinfo{year}{1977}\natexlab{}.
\newblock \showarticletitle{A Universal Algorithm for Sequential Data
  Compression}.
\newblock \bibinfo{journal}{\emph{IEEE Transactions on Information Theory}}
  \bibinfo{volume}{23}, \bibinfo{number}{3} (\bibinfo{year}{1977}),
  \bibinfo{pages}{337--343}.
\newblock


\end{thebibliography}

\end{document}